\documentclass[preprint,12pt]{elsarticle}
\usepackage{float}
\usepackage{amssymb}
\usepackage{amsmath}
\usepackage{subfig}
\usepackage{relsize}
\usepackage[linesnumbered,ruled,vlined]{algorithm2e}
\SetKwInput{KwInput}{Input}
\SetKwInput{KwOutput}{Output}
\journal{Ad Hoc Networks}

\begin{document}

\begin{frontmatter}



\title{Autonomous Vision-Aided UAV Positioning for Obstacle-Aware Wireless Connectivity}


\author{Kamran Shafafi} %% Author name

\affiliation{organization={INESC TEC, Faculdade de Engenharia, Universidade do Porto},%Department and Organization
            addressline={Rua Dr. Roberto Frias}, 
            city={Porto},
            postcode={4200-465}, 
            state={Porto},
            country={Portugal}}
            
\author{Manuel Ricardo} %% Author name

\author{Rui Campos} %% Author name

%% Abstract
\begin{abstract}
Unmanned Aerial Vehicles (UAVs) offer a promising solution for enhancing wireless connectivity and Quality of Service (QoS) in urban environments, acting as aerial Wi-Fi access points or cellular base stations. Their flexibility and rapid deployment capabilities make them suitable for addressing infrastructure gaps and traffic surges. However, optimizing UAV positions to maintain Line of Sight (LoS) links with ground User Equipment (UEs) remains challenging in obstacle-dense urban scenarios. This paper proposes VTOPA, a Vision-Aided Traffic- and Obstacle-Aware Positioning Algorithm that autonomously extracts environmental information~--~such as obstacles and UE locations~--~via computer vision and optimizes UAV positioning accordingly. The algorithm prioritizes LoS connectivity and dynamically adapts to user traffic demands in real time. Evaluated through simulations in ns-3, VTOPA achieves up to a 50\% increase in aggregate throughput and a 50\% reduction in delay, without compromising fairness, outperforming benchmark approaches in obstacle-rich environments.

\end{abstract}


\begin{highlights}
\item Vision-assisted UAV positioning techniques for 6G-enabled applications
\item Development of a traffic- and obstacle-aware algorithm for UAV positioning
\item Autonomous environmental perception enabling Line-of-Sight communications
\end{highlights}

%% Keywords

\end{frontmatter}

%%%%%%%%%%%%%%%%%%%%%%%%%%%%%%%%%%%%%%%%%%%%%%%%%%%%%%%%%%%%%%%%%
% INTRODUCTION
%%%%%%%%%%%%%%%%%%%%%%%%%%%%%%%%%%%%%%%%%%%%%%%%%%%%%%%%%%%%%%%%%
\section{Introduction} \label{Introduction-Section}

%%% CONTEXT
UAVs are increasingly being explored as aerial Wi-Fi access points (APs) or cellular base stations (BSs) to enhance wireless network coverage and capacity, particularly in dense and underserved urban environments \cite{zeng2016wireless, yanmaz2018drone}. Due to their mobility and flexibility, UAVs can be rapidly deployed to provide temporary or on-demand connectivity during peak traffic periods, emergency situations, or infrastructure outages. Several real-world deployments have demonstrated their potential to improve Quality of Service (QoS) and extend coverage in next-generation networks \cite{10.1007/978-3-031-57523-5_19}. However, leveraging UAVs as aerial access points introduces significant challenges, particularly in urban areas with complex landscapes and high building densities. One of the most critical issues is maintaining Line of Sight (LoS) connectivity between the UAV and ground UEs. Obstacles such as buildings can severely degrade signal quality, resulting in reduced throughput and increased latency. This makes UAV positioning not only a question of signal strength but also of obstacle avoidance.
% %%%%% FIGURE %%%%%
\begin{figure*}[t]
    \centering
    \includegraphics[width=1\textwidth]{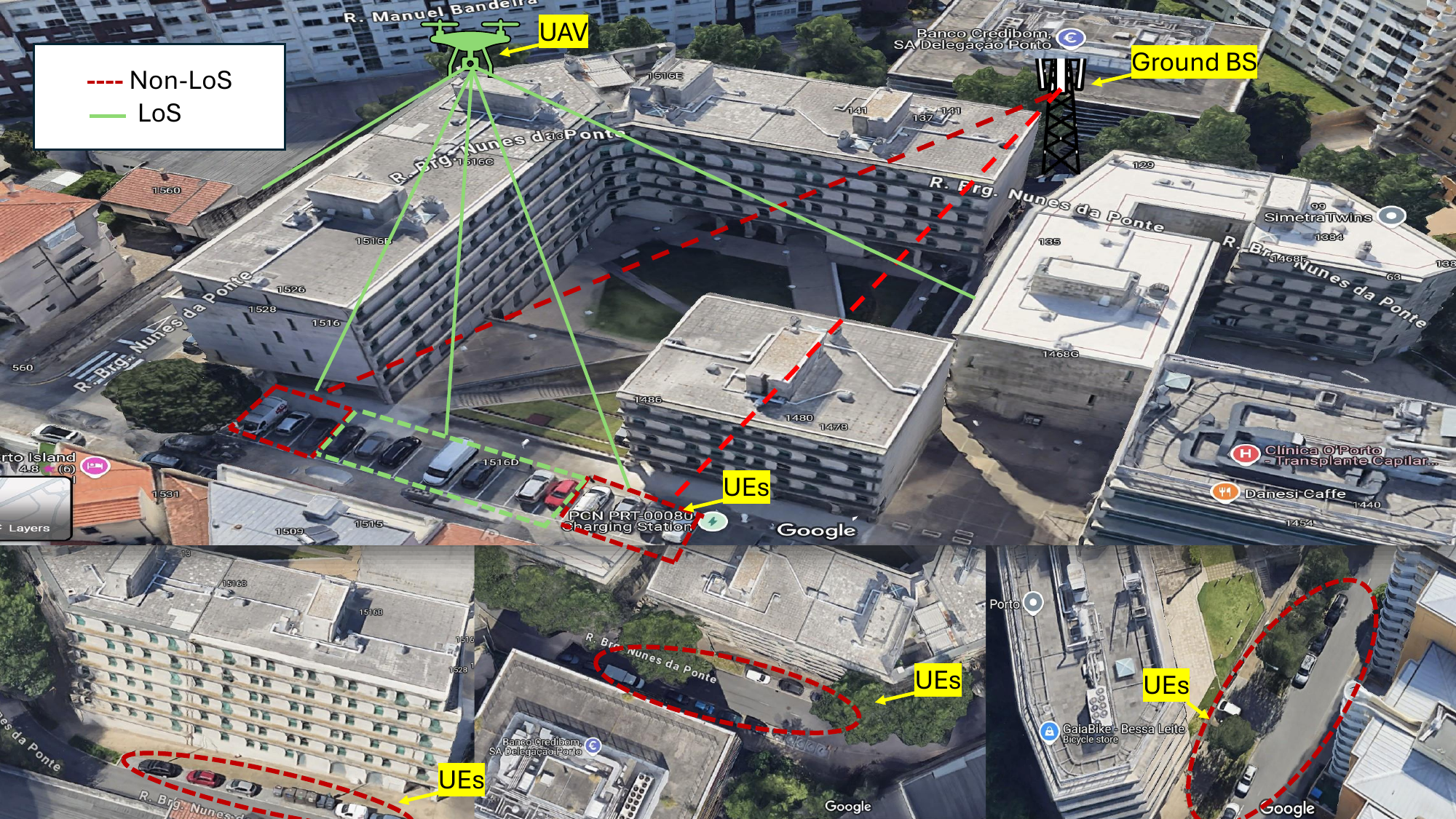} % Full page width
    \caption{Screenshots of different views of an urban environment in Porto, Portugal, showing a UAV positioned in 3D as an aerial AP to provide LoS wireless connectivity to UEs with the UAV connected to the ground BS.}
    \label{fig1}
\end{figure*}
% %%%%%%%%%%%%%%%%%%

Most existing approaches for UAV placement assume complete environmental knowledge or rely on offline simulations and planning \cite{8422376, 8038014}. These methods often neglect the dynamic nature of urban wireless traffic and the variability of user distributions. Moreover, few consider sensing and environmental awareness as part of the decision-making process. There is thus a need for UAV placement strategies that are both obstacle-aware and traffic-aware, while operating in real time without prior environmental knowledge \cite{shafafi2025MTOPA}. In \cite{10539557}, we proposed the Traffic- and Obstacle-aware UAV Positioning Algorithm (TOPA), which optimizes UAV positioning to maintain LoS connectivity with multiple UEs. However, it assumes predefined UE and building coordinates and does not account for complex urban environments with multiple obstacles of different shapes. In \cite{10792915, shafafi2025frameworkdevelopvalidaterlbased}, we presented the Reinforcement Learning-based Traffic and Obstacle-aware Positioning Algorithm (RLTOPA), which optimizes UAV positioning using Reinforcement Learning (RL) and considering multiple obstacles. However, while it improves obstacle awareness, it still does not address realistic building shapes and remains non-autonomous, relying on predefined UE and building coordinates.

To address this gap, we propose Vision-Aided Traffic- and Obstacle-Aware Positioning Algorithm (VTOPA), a novel UAV placement strategy that uses onboard cameras and computer vision to autonomously detect obstacles and localize ground users. As illustrated in Figure \ref{fig1}, the UAV processes visual input to map its environment and optimize its position, prioritizes maintaining LoS connectivity where possible, and dynamically adapts to user traffic demands in real time. The algorithm is implemented and evaluated using Network Simulator 3 (ns-3). Compared to state-of-the-art solutions, VTOPA improves aggregate throughput by up to 50\% and reduces packet delay by a similar margin while maintaining fairness across users. These results highlight the potential of combining vision-based perception with UAV mobility to enable smarter, adaptive aerial wireless networks in complex urban environments.

%%% ARTICLE CONTRIBUTIONS
The major contributions of this paper are:

\begin{itemize}
    \item \textbf{Vision-aided UAV positioning}: We propose a novel integration of vision and UAV-assisted wireless networks to autonomously detect UEs and obstacles and optimize UAV position;

    \item \textbf{Traffic- and obstacle-aware positioning algorithm}: We propose a Particle Swarm Optimization (PSO)-based algorithm that determines the UAV's optimal position. This method accounts for varying user traffic demands and urban obstacles while aiming to prioritize LoS connectivity where feasible between UEs and the UAV;
    
    \item \textbf{Autonomous environmental awareness}: In contrast to previous works, our approach eliminates the reliance on predefined UE and obstacle coordinates by autonomously extracting this information from the environment.

\end{itemize}

%%% ARTICLE STRUle broadbans liks to the ground users in CTURE
The rest of this article is organized as follows.
Section \ref{RelatedWork-Section} discusses the related work.
Section \ref{sec:systemmodel} presents the system model.
Section \ref{sec:formulation} formulates the problem.
Section \ref{Algorithm-Section} presents the fundamentals of the VTOPA in detail.
Section \ref{sec:Optimization Results} illustrates the optimal positions achieved by VTOPA across different scenarios, and discusses the evaluation results.
Section \ref{Conclusions-Section} draws the conclusions and points of the future work.

%%%%%%%%%%%%%%%%%%%%%%%%%%%%%%%%%%%%%%%%%%%%%%%%%%%%%%%%%%%%%%%%%
% RELATED WORK
%%%%%%%%%%%%%%%%%%%%%%%%%%%%%%%%%%%%%%%%%%%%%%%%%%%%%%%%%%%%%%%%%
\section{Related Work} \label{RelatedWork-Section}

UAVs have gained substantial attention in recent years as enablers of aerial wireless communication systems. Their mobility, low deployment cost, and on-demand operation capabilities make them particularly attractive for providing rapid connectivity in challenging scenarios such as natural disasters, large public events, and smart city deployments. These characteristics have motivated various research efforts focusing on integrating UAVs into existing cellular infrastructures to expand network coverage and enhance QoS.

Most studies on UAV placement and positioning assume free-space environments without considering physical obstructions. These works typically focus on positioning algorithms for UAVs acting as backhaul relays or access points within the Radio Access Network (RAN). For instance, \cite{8422376} proposed an efficient UAV-based backhaul network, while the authors of \cite{9448966} introduced a traffic-aware positioning algorithm tailored to aerial backhaul links. Other contributions focus on disaster-response and infrastructure-free deployments. In \cite{TianqiongChen}, a deployment strategy for post-disaster scenarios was proposed, whereas \cite{LIU2023103047,8038014, 7738405, 7572068} aimed to minimize the number of UAVs required while maintaining acceptable QoS and Quality of Experience (QoE) for UEs.

Recently, machine learning has been incorporated into UAV positioning algorithms. Deep learning approaches have been used to enhance network capacity in \cite{43-bab6dcd317da4e3d82375669cbb45023}, and reinforcement learning techniques such as Q-learning have been applied in \cite{23-8644345, 21-8377340, 22-COLONNESE2019101872} to optimize UAV positioning in dynamic environments. However, these studies predominantly consider free-space environments, which may not accurately reflect the complexity of real-world urban scenarios. 

Only a few works explicitly address UAV operations in obstacle-rich or urban environments. Most of these focus on modeling propagation losses and channel attenuation rather than proposing novel positioning strategies. For instance, \cite{8352733} proposed a propagation loss model for Super High-Frequency (SHF) frequencies in urban environments. In \cite{8450437}, the authors aimed to reduce the uplink latency between UEs and a single UAV. Similarly, \cite{s22030977} presented a propagation loss model to estimate LoS blockage probability for UAV base stations in urban areas. \cite{10199180} and \cite{10210623} investigated resource allocation in multi-UAV scenarios, optimizing UE-to-UAV associations to maximize overall throughput. Other works have focused on improving propagation modeling accuracy \cite{111114483593, 22227037248}, minimizing communication delays \cite{wu2021delay}, or analyzing LoS probability under different obstacle configurations \cite{Hayajneh20213DDO}.

In the domain of vision-aided UAV systems, most research has focused on autonomous flight tasks, such as landing, navigation, landing, obstacle avoidance, and trajectory recognition. For instance, \cite{7970402} employed computer vision UAV systems to enhance precision for quadrotor UAVs by mitigating GPS errors. The authors of \cite{10376356} introduced the first publicly available dataset for UAV self-positioning based on a dense sampling of low-altitude urban imagery. Further contributions include vision-based UAV navigation methods \cite{6852209}, and enhanced triangulation techniques for indoor UAV localization \cite{app14135646}. Despite the increasing integration of computer vision in UAV platforms, there remains a gap in applying these techniques to support real-time, obstacle-aware positioning solutions in UAV-assisted wireless networks. Most vision-based approaches are tailored to autonomous flight rather than communication optimization, and UAV positioning solutions for wireless networks rarely incorporate onboard sensing or perception capabilities. 

To the best of our knowledge, this paper presents the first vision-aided UAV positioning solution that explicitly targets joint obstacle-awareness and traffic-awareness for optimizing wireless coverage in urban environments. Our work integrates visual sensing with network-level decision-making, enabling UAVs to maintain LoS connectivity with UEs while dynamically adapting to spatial traffic demands.

\section{System Model}
\label{sec:systemmodel}

We consider a UAV-assisted aerial wireless communication system deployed over an urban area to provide wireless coverage for ground users in the presence of obstacles such as buildings. The system operates in a 3D space and consists of a single rotary-wing UAV acting as a mobile access point, multiple ground UEs, and various static urban obstacles. An overview of the system is illustrated in Fig. 1, where the UAV maintains wireless connectivity with UEs while ensuring LoS links. The system is composed of the following key elements:

\begin{itemize}
    \item \textbf{Unmanned Aerial Vehicle (UAV):} A single rotary-wing UAV, modeled by its 3D coordinate $P_u = (x_u, y_u, z_u)$, acts as a mobile access point. The UAV is equipped with a downward-facing RGB camera that captures aerial images of the venue. The UAV’s altitude $z_u$ is constrained by regulatory flight limits and the surrounding environment, such that $z^{Min} \leq z_u \leq z^{Max}$.

    \item \textbf{User Equipments (UEs):} Each UE represents a ground-based user statically located at position $P_i = (x_i, y_i, 0)$, where $i \in \{1, \dots, N\}$. The coordinates are extracted using visual detection from aerial imagery. UEs are assumed to be equipped with wireless communication interfaces.

    \item \textbf{Obstacles (Buildings):} The environment includes $M$ buildings, denoted as $\mathcal{B} = \{1,2,\ldots,M\}$. Each building is modeled as a 3D polygonal prism, characterized by a closed 2D polygonal footprint on the ground and extended vertically to a specified height. The base of the prism lies at ground level, while the top surface is flat and located at a defined elevation. The 3D structure is represented using two sets of vertices: the $bottom\_corners$, which define the ground-level footprint, and the $top\_corners$, which share the same horizontal coordinates but are vertically translated to the building’s height. The lateral surfaces are implicitly formed by connecting corresponding vertices between the top and bottom sets, enabling support for buildings with arbitrary polygonal footprints. This geometric model is utilized for LoS evaluation, where intersection tests are performed between the UAV-UE link and building volumes. The algorithm determines potential obstructions by combining 2D polygon intersection checks with vertical height comparisons, ensuring accurate detection of LoS blockages in complex urban environments.

    \item \textbf{Propagation Model:} Wireless propagation is modeled using the International Telecommunication Union (ITU) path loss models. Each air-to-ground link between the UAV and a $\mathrm{UE}_i$ is characterized by i) the 3D Euclidean distance $d_{u,i} = \|P_u - P_i\|$; ii) the presence or absence of LoS, based on geometric analysis; and iii) path loss and signal-to-noise ratio (SNR) determined according to the ITU-R P.1411 model \cite{ITU-RP1411}, considering antenna gains, transmit power, and noise level.
    
    \item \textbf{Central Processing Module (CPM):} A processing unit located at the network edge, in the cloud, or onboard the UAV. The CPM is responsible for: (i) collecting aerial visual data from the UAV camera; (ii) identifying UE and obstacle coordinates using computer vision techniques; and (iii) executing the VTOPA algorithm (detailed in Section~\ref{Algorithm-Section}) to determine a feasible UAV placement. The UAV then adjusts its location accordingly.

\end{itemize}
The described system aims to support real-time adaptive UAV deployment in urban environments, leveraging environment-aware sensing. This model currently considers static obstacles and user locations. Extensions to dynamic scenarios (e.g., moving vehicles or mobile users) are left for future work.

%%%%%%%%%%%%%%%%%%%%%%%%%%%%%%%%%%%%%%%%%%%%%%%%%%%%%%%%%%%%%%%%%
% Problem Formulation
%%%%%%%%%%%%%%%%%%%%%%%%%%%%%%%%%%%%%%%%%%%%%%%%%%%%%%%%%%%%%%%%%

\section{Problem Formulation}
\label{sec:formulation}

The system operates over discrete time slots denoted by \(\tau_k\), where \(k \in \mathbb{N}\). During each time slot \(\tau_k\), the positions of the UAV and all UEs are assumed to remain fixed. Let \(\mathcal{U} = \{0, 1, \ldots, N\}\) denote the set of nodes, with node \(0\) representing the UAV and nodes \(1\) to \(N\) representing the UEs. Each $\mathrm{UE}_i$ \(i \in \{1, \ldots, N\}\) is characterized by a traffic demand \(T_i(\tau_k) \in \mathbb{R}_+\) (in bit/s), and a fixed bandwidth \(W\) (in Hz), assumed identical for all UEs. Let \(P_u = (x_u, y_u, z_u) \in \mathbb{R}^3\) denote the UAV’s 3D position, and \(P_i = (x_i, y_i, z_i) \in \mathbb{R}^3\) denote the fixed position of UE \(i\). The Euclidean distance between them is defined as:

% \footnotesize
\begin{equation}
    \begin{aligned}
        d_{u,i} = \|P_u - P_i\| = \sqrt{(x_u - x_i)^2 + (y_u - y_i)^2 + (z_u - z_i)^2}
    \end{aligned}
    \label{eq:d}
\end{equation}

\normalsize

Each wireless link $\ell_i(\tau_k)$ between the UAV and $\mathrm{UE}_i$  may be subject to LoS or Non-Line-of-Sight (NLoS) conditions, modeled by a binary variable \( \text{LoS}_i \in \{0, 1\} \), where \( \text{LoS}_i = 1 \) indicates that a direct LoS path exists between $\mathrm{UE}_i$ and the UAV. The SNR at the UAV from $\mathrm{UE}_i$ is given by:

\begin{equation}
    \begin{aligned}
        SNR_i &= \frac{P_{T,i} \cdot G_T \cdot G_R \cdot \lambda^2}{(4\pi)^2 \cdot d_{u,i}^2 \cdot L_i \cdot N_0 \cdot W} \\
       \end{aligned}
       \label{snr}
\end{equation}

\noindent
where \( P_{T,i} \) is the transmit power of $\mathrm{UE}_i$, \( G_T \) and \( G_R \) are the antenna gains at the transmitter and receiver, \( \lambda = \frac{c}{f} \) is the wavelength corresponding to carrier frequency \( f \), \( N_0 \) is the noise power spectral density (W/Hz), and \( L_i \) is the LoS/NLoS propagation loss factor. The constant \( L_i \) is defined as:

\begin{equation}
    L_i =
    \begin{cases}
        1, & \text{if } LoS_i = 1 \\
        10^{\frac{L_{\text{NLoS}}}{10}}, & \text{if } LoS_i = 0
    \end{cases}
\end{equation}

\noindent
where \( L_{\text{NLoS}} = L_{obs} \) is the additional path loss (in dB) under NLoS conditions, determined according to the ITU-R P.1411 propagation model for urban environments \cite{ITU-RP1411}. This value accounts for empirically observed losses due to obstruction, shadowing, and diffraction (e.g., 20–30 dB empirically based on
environment density).

The capacity of the wireless link between $\mathrm{UE}_i$ and UAV is given by the Shannon-Hartley theorem:
\begin{equation}
    C_i(\tau_k) = W \cdot \log_2(1 + \text{SNR}_i)
    \label{c_i}
\end{equation} 
\noindent
% Throughput definition considering link capacity constraint
Since the actual data served to each UE cannot exceed the channel capacity, the achievable throughput at time slot $\tau_k$ is defined as:

\begin{equation}
    R_i(\tau_k) = T_i(\tau_k), \quad \text{where } T_i(\tau_k) \leq C_i(\tau_k)
    \label{eq:bitrate-definition}
\end{equation}

% UAV positioning problem
The UAV positioning optimization problem at time slot $\tau_k$ can be formulated as:
\begin{subequations}\label{objectives}
    \small
    \begin{align}
        \underset{P_u}{\text{maximize}} \quad & R(\tau_k) = \sum_{i=1}^{N} R_i(\tau_k)   \label{eq:objective-function1} \\[1ex]
        \text{subject to:} \quad 
        & C(\tau_k) = \sum_{i=1}^{N} C_i(\tau_k) \leq C^{\text{MAX}} \label{eq:constraint1} \\
        & 0 < T_i(\tau_k) \leq C_i(\tau_k), \quad \forall i \in \{1, \dots, N\} \label{eq:constraint2} \\
        & x^{\text{Min}} \leq x_u \leq x^{\text{Max}} \label{eq:constraint3} \\
        & y^{\text{Min}} \leq y_u \leq y^{\text{Max}} \label{eq:constraint4} \\
        & z^{\text{Min}} \leq z_u \leq z^{\text{Max}} \label{eq:constraint5}
    \end{align}
\end{subequations}

The constraint (\ref{eq:constraint1}) ensures that the total link capacity does not exceed the maximum channel capacity $C^{\text{MAX}}$. The constraint (\ref{eq:constraint2}) ensures that each UE’s traffic demand is feasible and accommodated, given the channel conditions. The final three constraints define the 3D deployment region for the UAV based on the venue size.

As the wireless medium is shared among all UEs, a contention-based Medium Access Control (MAC) protocol such as Carrier Sense Multiple Access with Collision Avoidance (CSMA/CA) is assumed to coordinate transmissions and avoid collisions. The optimization problem is non-convex due to the nonlinear dependence of SNR on UAV–UE distance and the logarithmic capacity model. These characteristics preclude the use of gradient-based solvers and motivate the application of heuristic approaches to efficiently explore the search space.

%%%%%%%%%%%%%%%%%%%%%%%%%%%%%%%%%%%%%%%%%%%%%%%%%%%%%%%%%%%%%%%%%
% VTOPA
%%%%%%%%%%%%%%%%%%%%%%%%%%%%%%%%%%%%%%%%%%%%%%%%%%%%%%%%%%%%%%%%%
\section{VTOPA Algorithm} \label{Algorithm-Section}

This section details the Vision-Aided Traffic- and Obstacle-Aware Positioning Algorithm (VTOPA), which integrates computer vision (CV) techniques to extract environmental information, including building geometries and UE positions. Based on the extracted data, PSO is employed to determine the optimal UAV placement, ensuring maximum throughput while prioritizing LoS connectivity to the UEs. The following subsections describe each phase of the algorithm.

\subsection{Environmental Information Extraction} \label{EnV-Algorithm-Section} 
VTOPA uses computer vision techniques to autonomously collect environmental information required for UAV positioning. Specifically, it extracts the 3D coordinates of building vertices and the locations of ground UEs. Buildings are modeled as static obstacles, and potential UEs are users/persons, vehicles, etc., detected in the images captured from the UAV.

\begin{figure*}[t]
	\centering
	\subfloat[West view.] {
		\includegraphics[width=0.45\linewidth, , height=5cm]{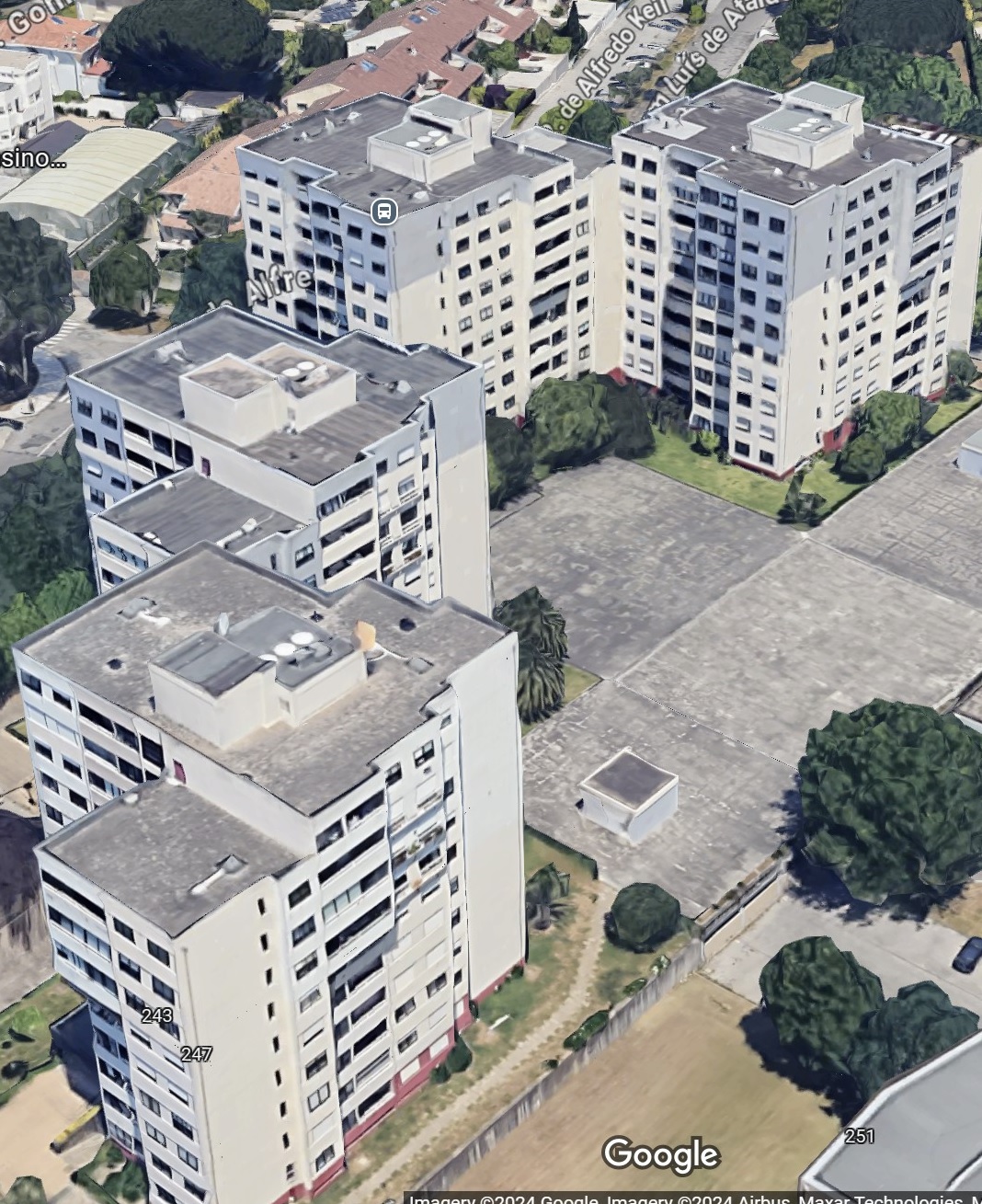}
		\label{fig2a}
	}
	\hfill
	\subfloat[Up to down view.] {
		\includegraphics[width=0.45\linewidth, height=5cm]{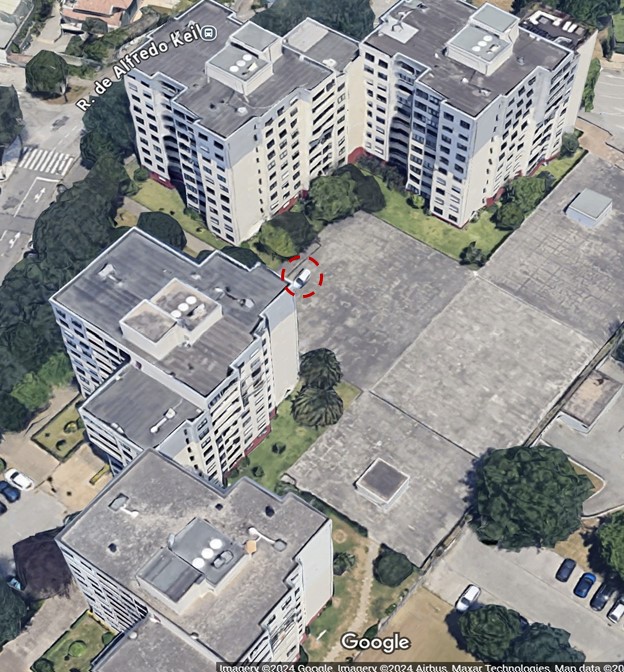}
		\label{fig2b}
	}
	\\
	\subfloat[North view.] {
		\includegraphics[width=0.45\linewidth, height=5cm]{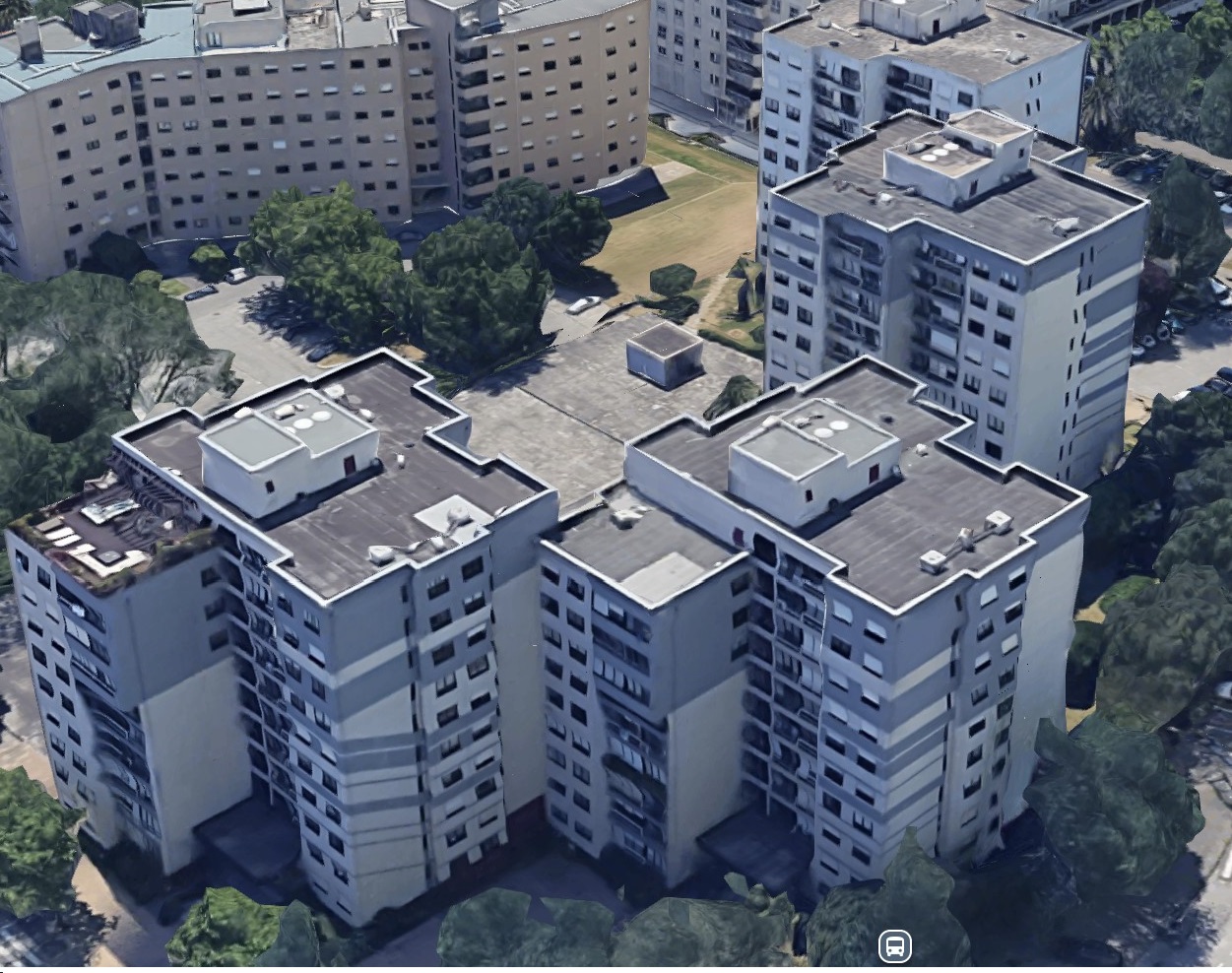}
		\label{fig2c}
	}
	\hfill
	\subfloat[South view.] {
		\includegraphics[width=0.45\linewidth, height=5cm]{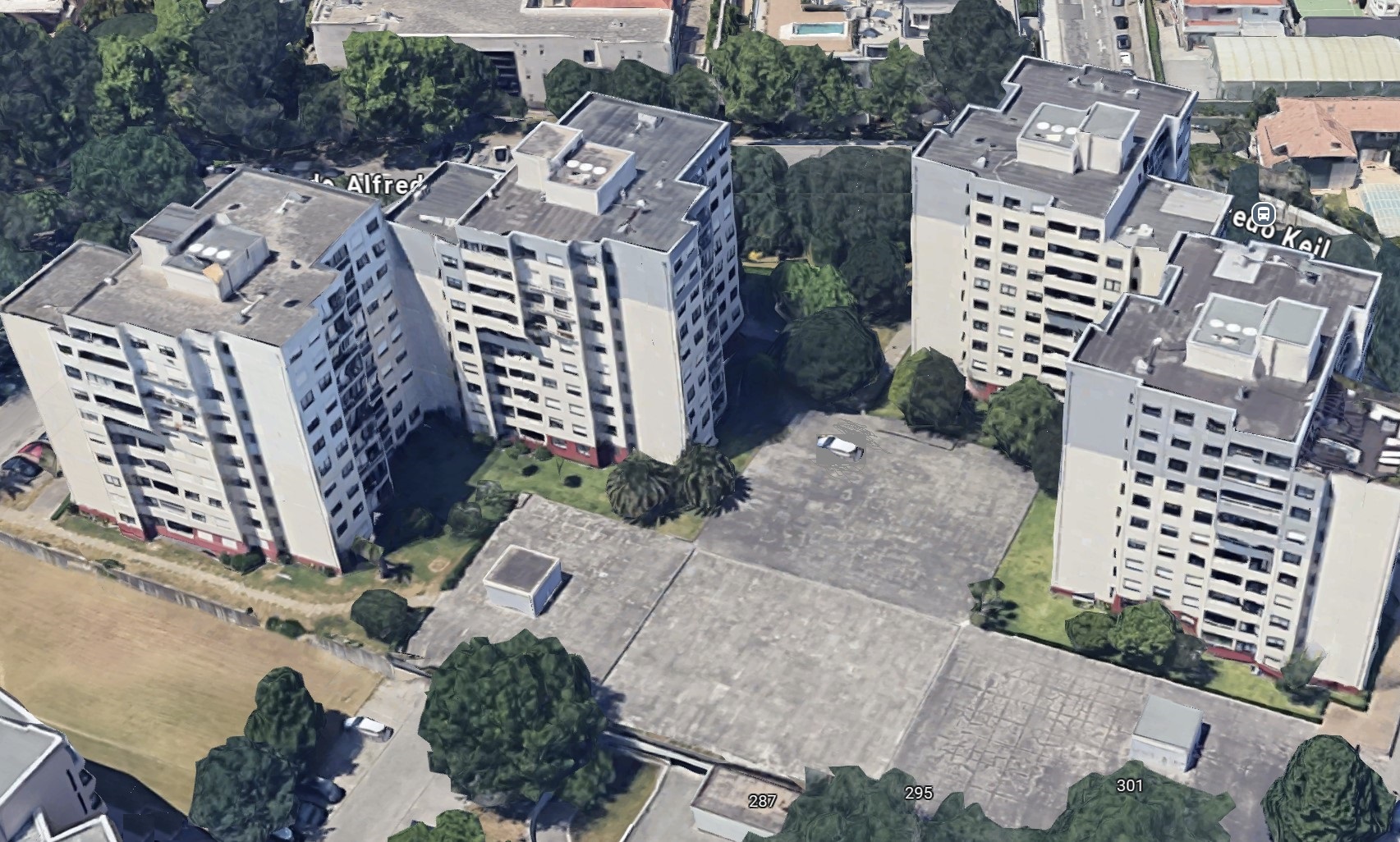}
		\label{fig2d}
	}
	\caption{Snapshots from Google Maps, capturing different views of a building area in Porto, Portugal, where the UAV is planned to be deployed.}
	\label{fig.2}
\end{figure*}

Autonomous detection can operate in two modes: real-time imagery and detection or offline processing using pre-captured images. In both modes, the system accounts for the angle of view by employing a multi-view analysis approach. This involves using at least four images from different viewpoints to estimate building structures and UE position accurately. This multi-view approach enhances the detection of occluded UEs and improves the estimation of building heights and footprints. In real-time adaptation, the CPM processes the images either onboard the UAV or at the network edge to extract environmental features on-the-fly. In this mode, the images are actively captured by the UAV during flight. In contrast, in the offline mode, pre-captured imagery~--~such as Google Maps snapshots of the target area~--~can be used as input for detection. These features are extracted ahead of time during pre-processing. Figure \ref{fig.2} illustrates this with example snapshots from an urban area in Porto, Portugal.

%%%%% FIGURE %%%%%

The UE detection process in VTOPA is powered by an automated object detection framework utilizing the You Only Look Once, version~8 (YOLOv8) deep learning architecture. The specific model, identified as ``satellite/6'', is deployed via the Roboflow platform~\cite{Roboflow40:online} and is tailored for object detection from satellite and drone imagery, ensuring effective performance for VTOPA's detection tasks. The detection script is implemented in Python, integrating the Roboflow Inference SDK for model deployment and OpenCV for image processing. The process begins with loading and preprocessing the input image using OpenCV to prepare it for analysis. Model inference is performed using Roboflow's HTTP client, with a confidence threshold set at 0.05 to filter detections. Post-processing extracts the bounding box coordinates for the detected vehicles, identifying four corners~--~top left, top right, bottom left, and bottom right.

The results are compiled into a JSON file containing detailed coordinate information for each vehicle and visualized in an annotated image with bounding boxes, labeled by car number and confidence score. The script leverages OpenCV for image annotation and NumPy for efficient coordinate mapping, producing a 2D representation of the environment. This implementation supports accurate detection in urban settings. In addition, pixel coordinates are converted into metric measurements by referencing average vehicle dimensions, facilitating a real-world interpretation of the data. The center of each car is considered as the coordinate of the UE.

\begin{figure}[t]
    \centering
    \begin{minipage}[t]{0.48\linewidth}
        \vspace{0pt} % Ensures top alignment
        \centering
        \includegraphics[width=\linewidth]{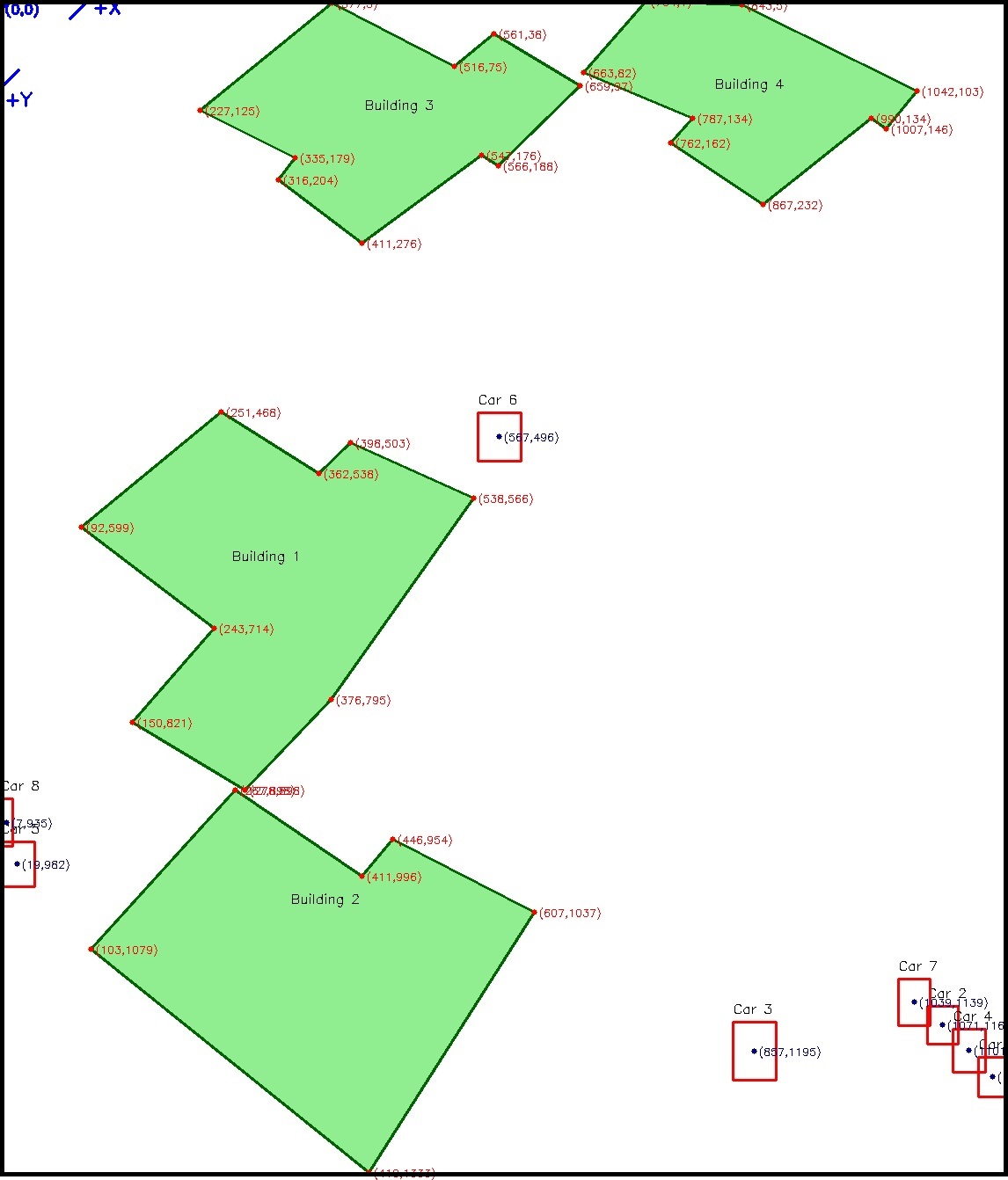}
        \caption{The 2D mapping of the desired scenario includes annotated buildings and cars, along with the coordinates of the building vertices and car positions.}
        \label{fig3}
    \end{minipage}
    \hfill
    \begin{minipage}[t]{0.48\linewidth}
        \vspace{0pt} % Ensures top alignment
        \centering
        \includegraphics[width=\linewidth, height =7.7 cm]{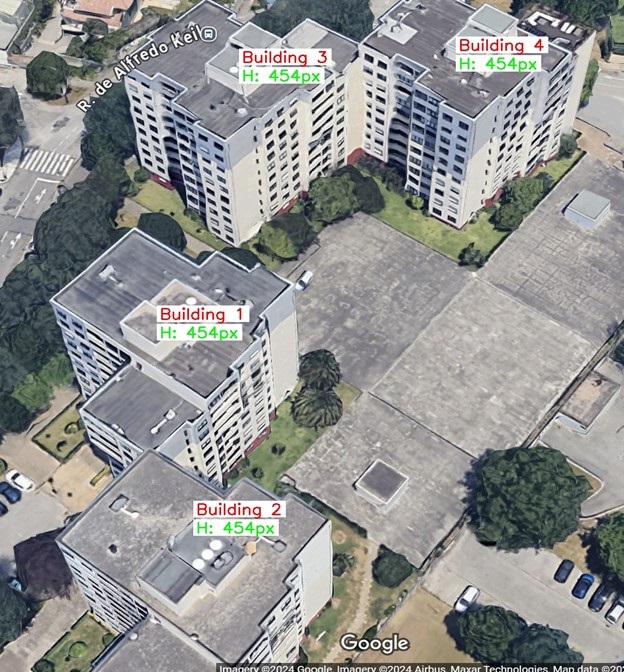}
        \caption{Annotated building heights for the desired scenario.}
        \label{fig4}
    \end{minipage}
\end{figure}

The multi-view analysis, utilizing four or more images with 45--90\textdegree\ angular separation, ensures robust detection by accounting for varying angles of view. For UE detection, YOLOv8, supported by OpenCV preprocessing and NumPy coordinate mapping, mitigates occlusion and perspective distortions to accurately identify UE positions. Similarly, for building footprint detection, OpenCV's contour detection and NumPy's coordinate aggregation combine multiple perspectives to precisely extract building vertices.

A 2D mapping system for roof building detection and coordinate extraction is also implemented using Python, integrating multiple CV libraries. The system utilizes OpenCV (cv2) for image processing and visualization. The process begins with the initial visualization of the detected vehicles on a white canvas, followed by an annotation system, and coordinates are efficiently mapped using NumPy arrays. This implementation combines OpenCV's precision with NumPy's computational efficiency, enabling accurate building capture in urban environments with intricate structures. Two sets of vertices, including $bottom\_corners$ that define the ground-level footprint, and $top\_corners$, are saved into a JSON file. Figure \ref{fig3} shows the resulting 2D map of the environment. As shown in Figure \ref{fig3}, the origin of the Cartesian coordinate system is located at the top-left corner of the photo, with \(Z=0\) indicating the ground level. All the environmental information required to proceed with the PSO-based positioning solution is prepared, including the 3D coordinates of the UEs, the vertices of the buildings, and the heights of the buildings.

We developed a measurement system for building height estimation, integrating edge detection with multi-view analysis. The system leverages OpenCV for image processing and NumPy for numerical computations. Using edge detection, images are converted to grayscale, and the $Sobel$ operator is applied to identify vertical edges, followed by normalization of the results. Next, vertical lines are detected using the $Hough Line Transform$, with the parameters tuned for building height estimation. The detected vertical lines are processed to measure building heights, incorporating multiple views for accuracy and filtering invalid measurements. Statistical processing combines these multi-view measurements, using median values to mitigate outliers and computing standard deviation for validation. The estimated height of the building is shown in Figure \ref{fig4}. The heights are then converted into metric units using the same method for the corner coordinates of the building. This robust edge detection-based approach effectively identifies and measures building vertical edges, ensuring reliable height estimation.

The extracted positions of the UEs, the vertices of the building, and its height are used as inputs for the PSO-based positioning algorithm, described in the next section.

%%%%% FIGURE %%%%%
\begin{figure}[t]
    \centering
    \includegraphics[width=0.8\linewidth]{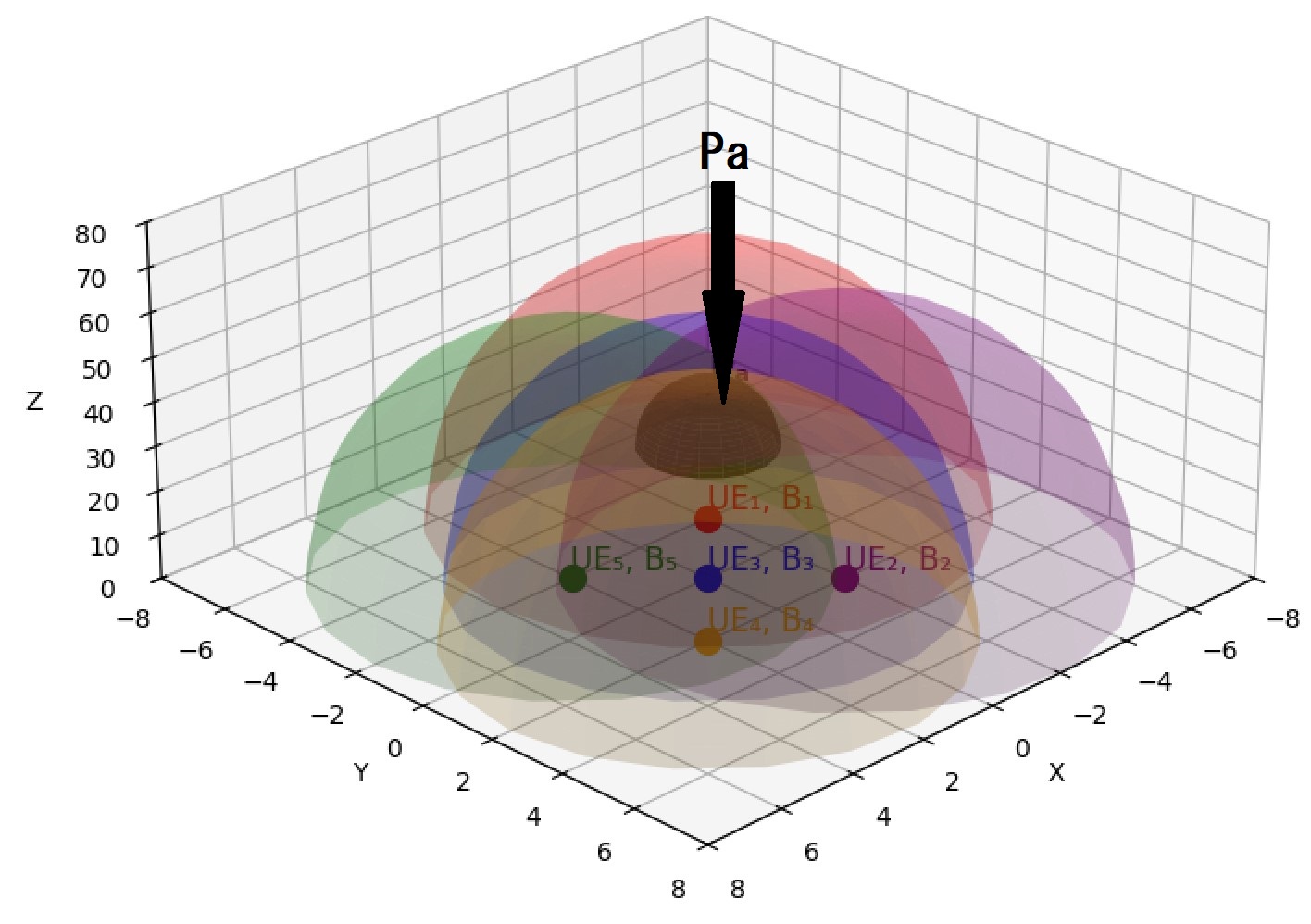}
    \caption{Illustration of five UEs with varying traffic demands. The feasible UAV positioning region $P_a$ is defined as the intersection volume of spheres, each centered at a UE and bounded by the maximum allowed distance between each UE and the UAV.}
    \label{fig5}
\end{figure}
%%%%%%%%%%%%%%%%%%
\subsection{PSO-based UAV Positioning} \label{PSO-Algorithm-Section}

This section outlines the underlying principles and rationales of the PSO-based UAV positioning strategy. Initially, a Modulation and Coding Scheme (MCS)~\cite{MCSTable50:online} is employed to identify the minimum required SNR between the UAV and each user equipment $\mathrm{UE}_i$, denoted as \( \text{SNR}_{\text{req},i} \), that can satisfy the traffic demand \( T_i(\tau_k)\). This requirement is then applied to each communication link \( \ell_i(\tau_k) \) (\(\triangleright\)~lines~1-2, Algorithm \ref{algorithm}). Taking into account \( \text{SNR}_{\text{req},i} \), the maximum allowed distance between each $\mathrm{UE}_i$ and the UAV to meet the traffic demand \( T_i(\tau_k) \) is defined as:

\begin{equation}
    d_{u_{\text{max}},i} = \sqrt{ \frac{P_{T,i} \cdot G_T \cdot G_R \cdot \lambda^2}{(4\pi)^2 \cdot L_i \cdot N_0 \cdot W \cdot \text{SNR}_{\text{req},i}} }
    \label{dui_max}
\end{equation}
\noindent
referring to Equation~\eqref{snr} (\(\triangleright\)~line~3, Algorithm \ref{algorithm}). This expression defines a spherical region of radius \( d_{u_{\text{max}},i} \) around each $\mathrm{UE}_i$, centered at \( P_i \), within which the UAV must be located to ensure the traffic demand of each $\mathrm{UE}_i$ is satisfied. Thus, the condition becomes:

\begin{equation}
    d_{u,i} \leqslant d_{u_{\text{max}},i} \quad\quad\quad \forall i \in \{1, \dots, N\}
    \label{dui_constraint}
\end{equation}

In the case of multiple UEs, multiple spheres are defined, each centered around its $\mathrm{UE}_i$. To meet the traffic demand of all UEs, the UAV must be positioned within the intersection of all these spheres, denoted as the position volume \(P_a \in \mathbb{R}^3\). Figure~\ref{fig5} illustrates a scenario involving five UEs with different traffic demands. The figure highlights the corresponding spheres centered at $P_i$, representing their respective \(d_{u_{\text{max}},i}\), and the intersection of these spheres, \(P_a\). However, the intersection between them may not always exist. In large-scale scenarios where numerous UEs are widely distributed with varying traffic demands~--~and consequently, varying radii~--~finding a single intersection that satisfies all UEs becomes challenging.

\begin{algorithm}
\scriptsize
\DontPrintSemicolon
\caption{PSO-based UAV Positioning}
\label{algorithm}
\KwInput{
    UEs 3D position \(P_i\);\quad
    obstacle vertices coordinates and height;\quad
    Traffic demand \(T_i\) (in \text{bit/s});\quad
    antenna gain  \(G_T\) and \(G_R\) ;\quad
    Noise floor power \(P_N\) (in \text{dBm});\quad
    Frequency \(f\);\quad
    Transmission power \(P_{T,i}\) (in \text{dBm});\quad
    Bandwidth $W$ (in Hz);\quad
    NLoS path loss \(L_\text{obs}\) (in \text{dB})
}

\tcc{\textbf{define a spherical region for each $\mathrm{UE}_i$}}
\For{\(i = 1; i \leq N; i \leftarrow i + 1\)}{
    Compute $\text{SNR}_{\text{req},i}$ according to \(T_i\), using MCS table\;
    Compute $d_{u_{\text{max}},i} = \sqrt{ \frac{P_{T,i} \cdot G_T \cdot G_R \cdot \lambda^2}{(4\pi)^2 \cdot L_i \cdot N_0 \cdot W \cdot \text{SNR}_{\text{req},i}} }$\;
}
\textbf{Identifying all possible intersections between spheres }\;
\textbf{Select the intersection with the maximum number of associated UEs as $P_a$}\;
\textbf{Discretize $P_a$ with 1.0\,m precision to define search space}

\tcc{\textbf{Filter Valid Positions from \(P_a\)}}
\For{each position \(P_u \in P_a\)}{
    \If{\(P_{u,x} \in [x^{\text{Min}}, x^{\text{Max}}]\) \textbf{and} \(P_{u,y} \in [y^{\text{Min}}, y^{\text{Max}}]\) \textbf{and} \(P_{u,z} \in [z^{\text{Min}}, z^{\text{Max}}]\)}{
        Add \(P_u\) to \textbf{ValidPositions}\;
        \tcc{$z^{\text{Min}}$ is the height of tallest builing}
    }
}
\tcc{\textbf{Initialize PSO Parameters }}
particles = 30, MaxIterations =100, \(w = 0.7\),\(c_1 = c_2 = 1.5\)\;
\tcc{\textbf{Initialize Particles}}
Set particle positions from \textbf{ValidPositions}\;
\(Iteration = 0\), \(maxR = 0\), \(P_{optimal} = \emptyset\)\;

\While{\(Iteration < MaxIterations\)}{
    \tcc{Evaluate fitness for all particles}
    \For{each particle}{
        \(P_u \gets\) current position of particle\;
        \(R = 0\)\;
        
        \For{\(i = 1; i \leq N; i \leftarrow i + 1\)}{
            Compute \(d_{u,i} = \sqrt{(x_u - x_i)^2 + (y_u - y_i)^2 + (z_u - z_i)^2}\)\;
            \hspace{1cm}\;
            \tcc{\textbf{Assess LoS link}}
            \If{\(P_u\) has LoS with $\mathrm{UE}_i$}{
                \(L_i = 1\)\;
                }
            \Else{
                \(L_i = 10^{\frac{L_{\text{NLoS}}}{10}}\)\;
                \( L_{\text{NLoS}} = L_{obs}\)\;
                \tcc{ $L_{obs}$ is the additional loss due to shadowing or buildings}
            }
            Compute $\text{SNR}_i$ from Equation \ref{snr}\;
            Compute $ C_i(\tau_k)$ from Equation \ref{c_i}\;
            $ R_i(\tau_k) = \min\left( T_i(\tau_k), C_i(\tau_k)\right))$\;
            \(R \gets R + R_i\)\;
        }
        
        \(fitness \gets R\)\;
        \tcc{\textbf{Update particle's best position if \(R\) improves}}
        \If{\(R > maxR\)}{
            \(P_{optimal} \gets \{P_u\}\)\;
            \(maxR \gets R\)\;
        }
        
       }
    
    \tcc{Update particle positions for next iteration}
    \For{each particle}{
        \tcc{\textbf{Compute velocity:}}
        \(v \gets w \cdot v + c_1 \cdot r_1 \cdot (p_{best} - P_u) + c_2 \cdot r_2 \cdot (g_{best} - P_u)\)\;
        \tcc{\textbf{Update target position:}}         
        \(P_u^{\text{target}} \gets P_u + v\)\;
        \textbf{select closest position in ValidPositions}\;
    }    
    \(Iteration \gets Iteration + 1\)\;
}
\KwOutput{\textbf{${P_{optimal}}$}}
\end{algorithm}

To address the challenge of ensuring coverage for the maximum number of UEs with a single UAV, we begin the process by identifying all possible intersections between spheres. Then, the intersection with the maximum number of associated UEs among all potential intersections is selected as $P_a$ (\(\triangleright\)~lines~4-5, Algorithm \ref{algorithm}). All feasible positions within $P_a$ are defined as the search space for the optimization process. By discretizing $P_a$ with 1.0 m precision, all possible UAV positions are considered to accommodate traffic demand and maximize throughput for the largest possible number of UEs using a single UAV. This process involves verifying which positions lie within $P_a$ and are valid for UAV placement~--~that is, positions above the highest buildings and within the venue boundaries ($\triangleright$~lines~6–9, Algorithm~\ref{algorithm}).

Although all candidate positions within $P_a$ satisfy the basic traffic requirements of their associated UEs, not all ensure LoS connectivity due to environmental obstacles such as buildings. In high-frequency communication scenarios, LoS connectivity is essential for establishing reliable, high-capacity links. While NLoS links can still support communication, they typically do so with significantly reduced throughput. To account for these environmental constraints, the VTOPA algorithm evaluates each candidate UAV position within $P_a$ by performing a geometric LoS check with respect to all UEs. This check involves constructing virtual straight-line segments between each candidate UAV position and each UE. These segments are tested for intersection with building surfaces using the Möller–Trumbore ray–triangle intersection algorithm~\cite{moller1997fast}, which enables accurate detection of physical obstructions. Based on the intersection results, each link is classified as either LoS or NLoS. This classification is then used to compute the individual UE throughputs $R_i(\tau_k)$, which collectively define the total effective throughput $R(\tau_k)$. This ensures that the optimization process captures realistic propagation conditions and environmental occlusions.

The PSO procedure iteratively updates each particle’s position in the search space to maximize the aggregate throughput $R(\tau_k)$. The fitness of each particle is evaluated based on the achieved throughput, and its position is updated using standard PSO dynamics. These dynamics include three components: an inertia term (reflecting the current velocity), a cognitive component (which attracts the particle toward its own best-known position, $p_\text{best}$), and a social component (which attracts the particle toward the global best-known position, $g_\text{best}$). The updated target position $P_u^\text{target}$ is computed using the PSO equations and then projected onto the nearest discrete valid location within the defined positioning area $P_a$ ($\triangleright$~lines~10–37, Algorithm~\ref{algorithm}).

The fitness function directly incorporates the LoS evaluation to ensure path obstructions are accurately modeled. Throughout the optimization process, the algorithm tracks all positions that yield the highest throughput, storing them in the set $P_\text{optimal}$. The PSO iterations continue until either the convergence criteria are met or the predefined maximum number of iterations is reached. The final outcome is a set of UAV positions that optimize network throughput while jointly considering LoS connectivity, spatial constraints, and the traffic demand distribution across UEs. By integrating environmental awareness into the optimization framework, VTOPA significantly enhances QoS in complex, obstacle-rich urban environments, as summarized in Algorithm~\ref{algorithm}.

%%%%%%%%%%%%%%%%%%%%%%%%%%%%%%%%%%%%%%%%%%%%%%%%%%%%%%%%%%%%%%%%%
% Optimization Results
%%%%%%%%%%%%%%%%%%%%%%%%%%%%%%%%%%%%%%%%%%%%%%%%%%%%%%%%%%%%%%%%%
\begin{table}[ht!]
    \centering
    \footnotesize
    \caption{\textbf{Summary of PSO algorithm parameters.}}
    \label{tab1}
    \setlength{\tabcolsep}{3pt}
    \begin{tabular}{p{0.4\textwidth} p{0.6\textwidth}}
        \hline
        \multicolumn{2}{c}{PSO Algorithm Parameters} \\  
        \hline    
        Number of particles & 30 $^{\mathrm{a}}$ \\
        Maximum iterations & 100 $^{\mathrm{b}}$ \\
        Position precision & 1.0 m $^{\mathrm{c}}$ \\
        Inertia weight (w) & 0.7 $^{\mathrm{d}}$ \\
        Cognitive coefficient (c1) & 1.5 $^{\mathrm{e}}$ \\
        Social coefficient (c2) & 1.5 $^{\mathrm{e}}$ \\
        Initial position space & Valid positions from $P_a$ \\
        Fitness function & Aggregate Throughput (R) \\
        Position constraints & UE-based boundaries with 2m margin \\
        Height constraints & Building height to 100m $^{\mathrm{f}}$ \\
        Early stopping & After 10 iterations if perfect solution \\
        Position selection & Closest valid position from $P_a$ \\
        \hline
        \multicolumn{2}{p{\textwidth}}{$^{\mathrm{a}}$ \footnotesize Standard PSO population size balancing exploration and computational cost} \\
        \multicolumn{2}{p{\textwidth}}{$^{\mathrm{b}}$ \footnotesize Sufficient iterations for convergence while maintaining efficiency} \\
        \multicolumn{2}{p{\textwidth}}{$^{\mathrm{c}}$ \footnotesize Practical precision for UAV positioning in real-world scenarios} \\
        \multicolumn{2}{p{\textwidth}}{$^{\mathrm{d}}$ \footnotesize Balances exploration and exploitation in particle movement} \\
        \multicolumn{2}{p{\textwidth}}{$^{\mathrm{e}}$ \footnotesize Standard PSO coefficients for personal and social learning} \\
        \multicolumn{2}{p{\textwidth}}{$^{\mathrm{f}}$ \footnotesize Maximum UAV positioning altitude} \\
        \hline
    \end{tabular}
\end{table}

\section{VTOPA Evaluation}\label{sec:Optimization Results}

This section presents the evaluation of VTOPA. In Subsection~\ref{sec:Validation of PSO-based positioning}, we validate the proposed PSO-based approach across three use cases with varying traffic demands and UE distributions. The analysis focuses on determining the optimal UAV positions and their ability to maintain LoS connectivity while satisfying throughput requirements. In Subsection~\ref{sec:evaluation Results}, we assess the performance of VTOPA through ns-3 simulations, comparing its aggregate throughput and mean delay against the RL-based approach (RLTOPA) proposed in \cite{10792915}, thereby highlighting its effectiveness and efficiency across diverse scenarios.

\subsection{Validation of PSO-based positioning}\label{sec:Validation of PSO-based positioning} 
Table \ref{tab1} presents the key parameters used in the PSO implementation. This PSO algorithm utilizes 30 particles to explore the solution space over a maximum of 100 iterations. The particles' movement is guided by standard PSO coefficients, including an inertia weight of 0.7 and cognitive and social coefficients of 1.5, to balance exploration and exploitation. The algorithm operates within position constraints derived from UE locations and building heights, ensuring realistic UAV placement. Positions are sampled with a precision of 1.0 meter to reflect practical UAV positioning capabilities.

VTOPA was tested across various scenarios with different traffic demands and number of UEs. However, without loss of generality, the scenario shown in Figure \ref{fig.2} is considered in this section as a straightforward example to assess the performance of VTOPA. It is worth noting that VTOPA can be applied to different real-world scenarios with different number of UEs and buildings, regardless of their shape or type. As shown in Figure \ref{fig3}, eight UEs are distributed within the venue, and three buildings are detected. We evaluate three different use cases applied to this scenario. 

\textbf{Use Case A} represents a homogeneous use case where all UEs have the same traffic demand of 58.5 Mbit/s, corresponding to the IEEE 802.11ac MCS index 0. In \textbf{Use Case B}, a heterogeneous setting is considered where \((B_1 = B_2 = B_3 = B_4) = 2 \times (B_5 = B_6 = B_7 = B_8)\). This assumes traffic demands of 117 Mbit/s, associated with the IEEE 802.11ac MCS index 1, and 58.5 Mbit/s, associated with the MCS index 0, respectively. \textbf{Use Case C} presents another heterogeneous setting where each UE has a different traffic demand. In this setting, the traffic demands associated with IEEE 802.11ac MCS indices 0 through 7 are applied to each UE, respectively.

\begin{figure}[t]
    \centering
    \begin{minipage}[t]{0.48\linewidth}
        \centering
        \includegraphics[width=\linewidth]{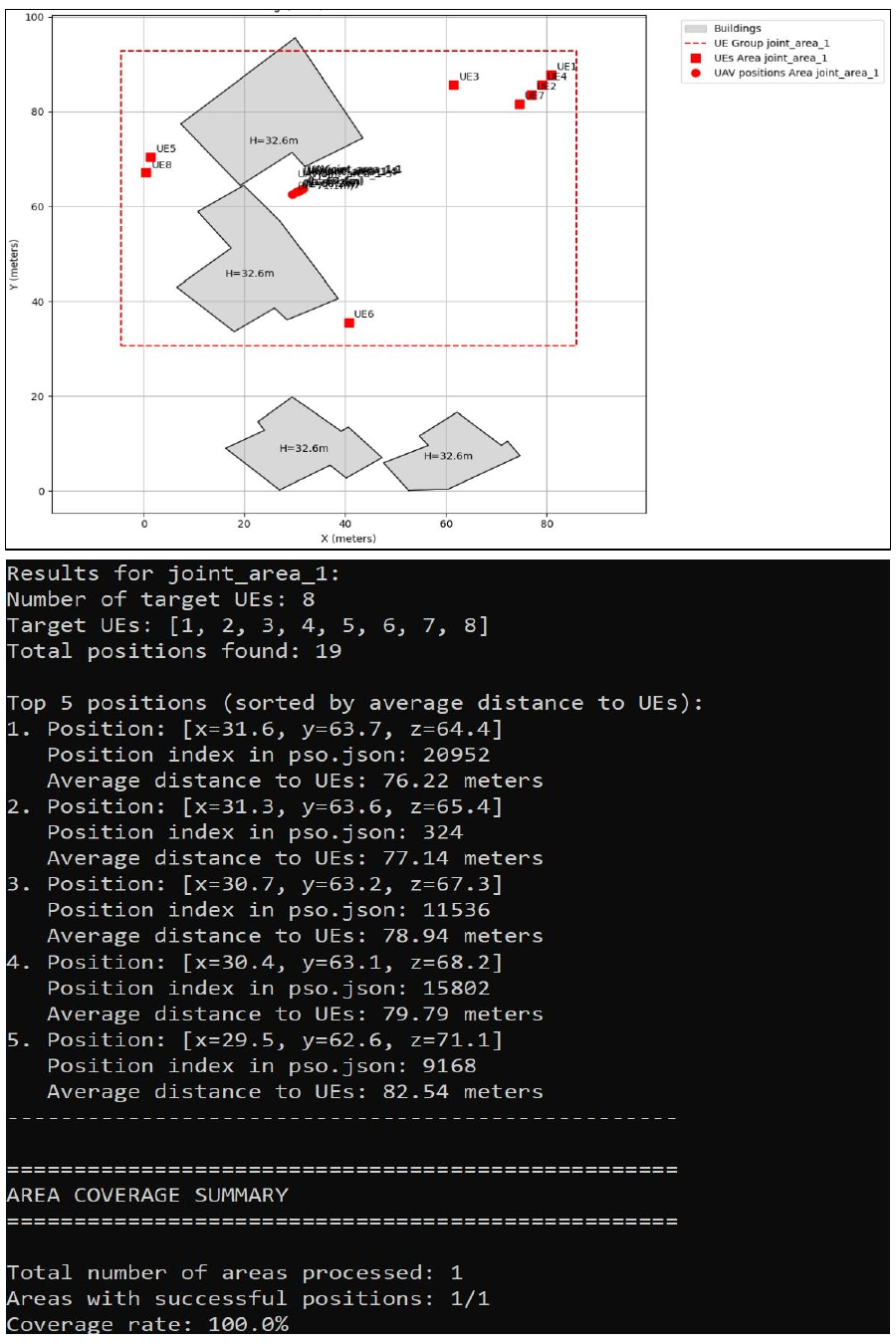}
        \caption{Use Case A with a traffic demand of 58.5 Mbit/s. VTOPA presented 19 optimal positions where the UAV maintains the LoS connection with all UEs and meets traffic demands.}
        \label{fig6}
    \end{minipage}
    \hfill
    \begin{minipage}[t]{0.48\linewidth}
        \centering
        \includegraphics[width=\linewidth]{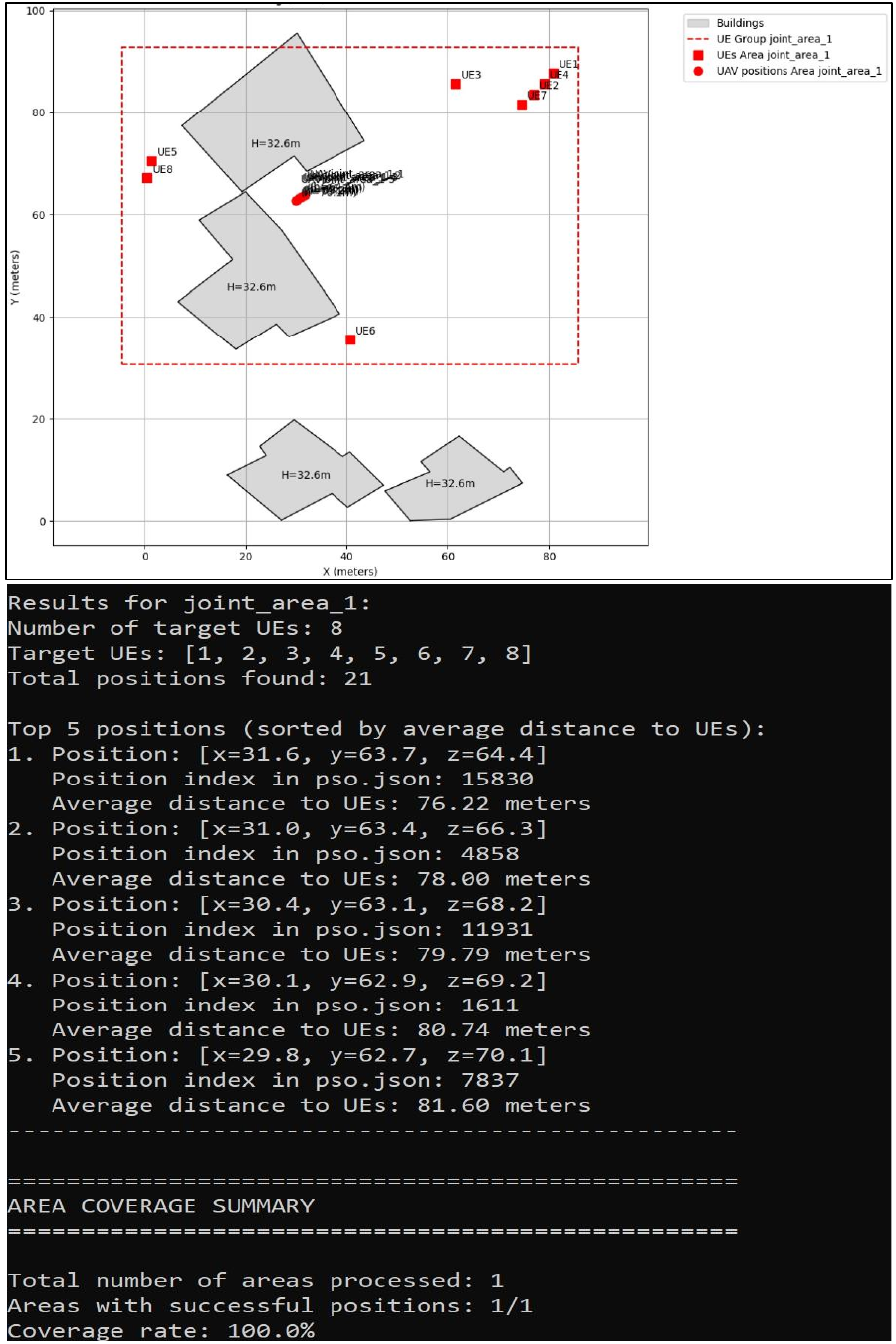}
        \caption{Use Case B features traffic demands of \((B_1 = B_2 = B_3 = B_4) = 2 \times (B_5 = B_6 = B_7 = B_8)\) associated with MCS index 0 to 1, respectively.}
        \label{fig7}
    \end{minipage}
\end{figure}

Figure \ref{fig6} shows the optimal position for Use Case A. The algorithm successfully identified an intersection, \(P_a\), associated with all UEs, within the red dashed rectangle. VTOPA presents a total of nineteen optimal positions that provide LoS to all UEs and meet the traffic demand of the UEs. The results show the five best UAV positions ordered by throughput from highest to lowest and the average distance of the UAV with associated UEs, where the first one is considered as $P_{optimal}$.

%%%%% FIGURE %%%%%
\begin{figure}[!t]
    \centering
    \includegraphics[width=0.6\linewidth]{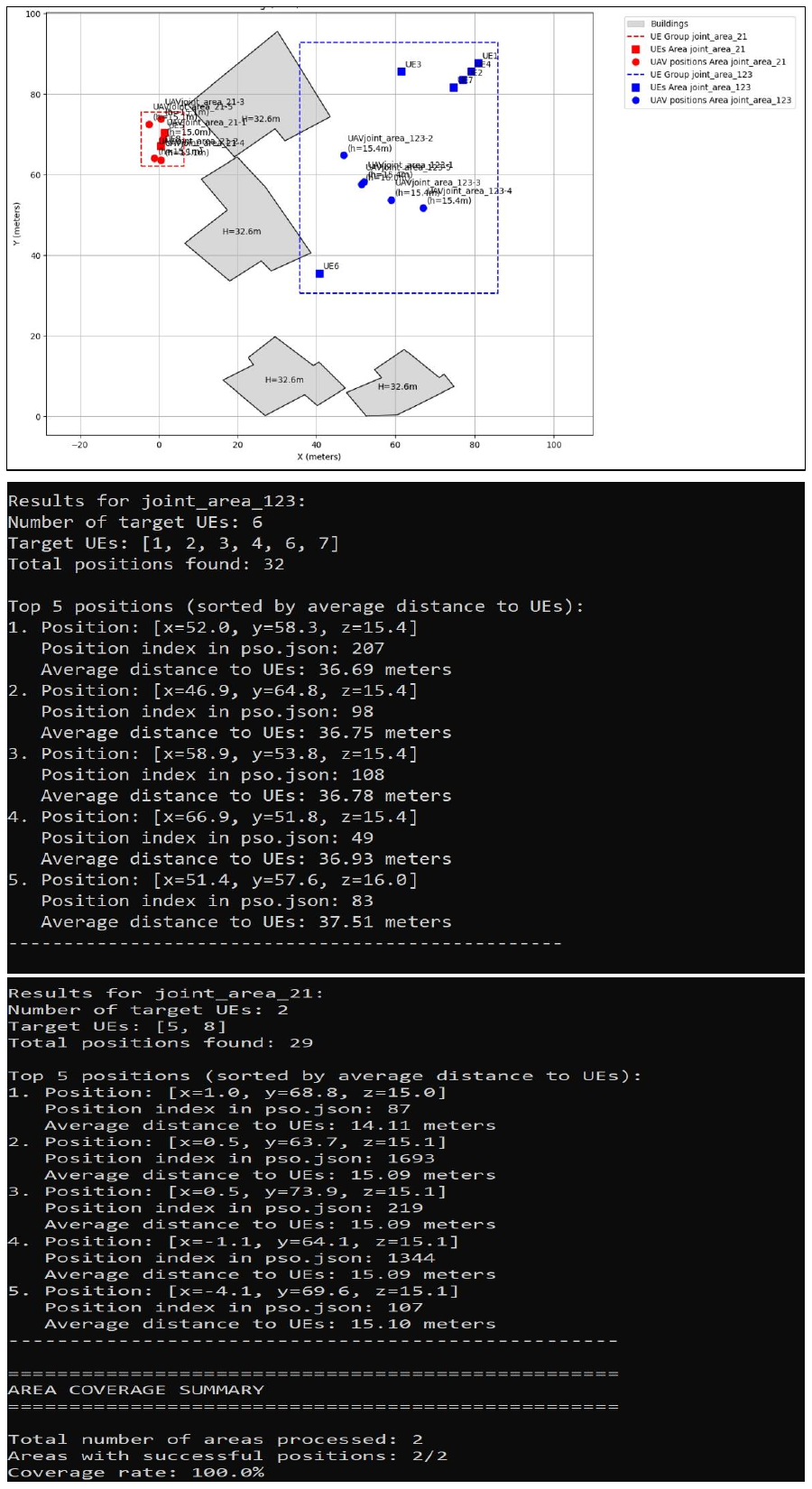}
    \caption{Use Case C all UEs with different traffic demands associated with MCS index 0 to 8, respectively.}
    \label{fig8}
\end{figure}
%%%%%%%%%%%%%%%%%%

Figure \ref{fig7} illustrates the optimal position achieved by VTOPA for Use Case B. As mentioned previously, this use case aimed to evaluate the performance of VTOPA in a heterogeneous setting where 50\% of UEs request twice the data rate of others. In this setting, the algorithm identified an intersection that satisfies the traffic demands of all UEs (red dashed rectangle). VTOPA found twenty-one positions that simultaneously meet the traffic demands and maintain LoS. 

Figure \ref{fig8} refers to Use Case C, a worst-case scenario that can arise in positioning problems. Two UEs request traffic demands exceeding 500 Mbit/s, associated with MCS indices 7 and 8. Higher traffic demands require higher \(SNR_i\), which consequently decreases \(d_{max,i}\) and creates smaller spheres around the desired UEs. In such cases, finding a unique intersection among all UEs becomes quite challenging. The algorithm, in this setting, covered all UEs in the venue by combining two intersection areas: one associated with the two high-demand UEs (red dashed rectangle) and another with six UEs (blue dashed rectangle). Since this paper focuses on a single UAV solution, the intersection area associated with more UEs is selected for UAV positioning. As illustrated in Figure \ref{fig8}, VTOPA provides optimal positions for both areas; however, the area including more UEs is ultimately chosen for the single UAV deployment.

Use Case C demonstrated that even in complex scenarios, VTOPA is capable of positioning the UAV in LoS with all UEs, similar to Use Case A and Use Case B. However, in some cases, it may be impossible to accommodate the traffic demands of all UEs using a single UAV. In such cases, multiple UAVs must be deployed, which is left for future work.

%%%%%%%%%%%%%%%%%%%%%%%%%%%%%%%%%%%%%%%%%%%%%%%%%%%%%%%%%%%%%%%%
% Evaluation Results
%%%%%%%%%%%%%%%%%%%%%%%%%%%%%%%%%%%%%%%%%%%%%%%%%%%%%%%%%%%%%%%%%
\subsection{Performance Evaluation of VTOPA}
\label{sec:evaluation Results}

\begin{table}[t]
    \centering
    \footnotesize
    \caption{\textbf{Simulation configuration of VTOPA.}}
    \label{tab2}
    \setlength{\tabcolsep}{3pt} % Reduce cell padding
    \begin{tabular}{p{0.55\textwidth} p{0.4\textwidth}}
        \hline        
        \multicolumn{2}{c}{ns-3.38 simulation parameters} \\
        \hline
        $S_{venue}$ & 100 m  \\        
        $N$ & variable with the scenario \\
        $f$ & 5250 MHz \\
        $M$ & variable with the scenario \\
        Guard Interval $(GI)$ & 800 $ns$ \\
        Wi-Fi channel & 50 \\
        Wi-Fi Standard & IEEE 802.11ac  \\
        Channel Bandwidth & 20 MHz  \\
        Antenna Gain & 0 dBi  \\
        Tx power & 20 $dBm$ \\
        Noise floor power & -85 $dBm$ \\
        LoS Propagation Loss Model & ItuR1411Los \\
        NLoS Propagation Loss Model & ItuR1411NlosOverRooftop \\
        Remote Station Manager mechanism & IdealWifiManager  \\
        Application Traffic & UDP constant bitrate \\
        UDP Data Rate $(T_i)$ & variable based on $MCS_i$ \\
        Packet Size & 1400 bytes  \\        
        \hline
    \end{tabular}
\end{table}

\begin{figure}[ht!]
    \centering
    \subfloat[Aggregate throughput (Mbit/s)]{\includegraphics[width=0.48\linewidth, height=6cm]{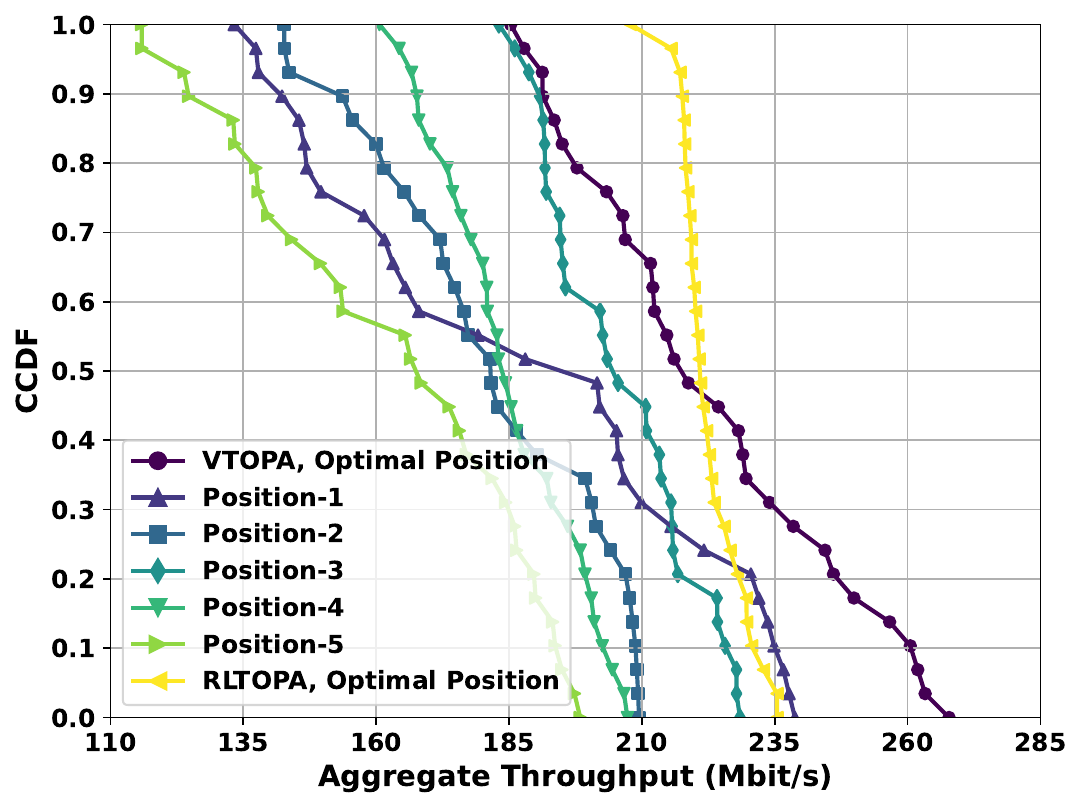}\label{fig9a}}
    \hfill
    \subfloat[Delay (ms)]{\includegraphics[width=0.48\linewidth, height=6.3cm]{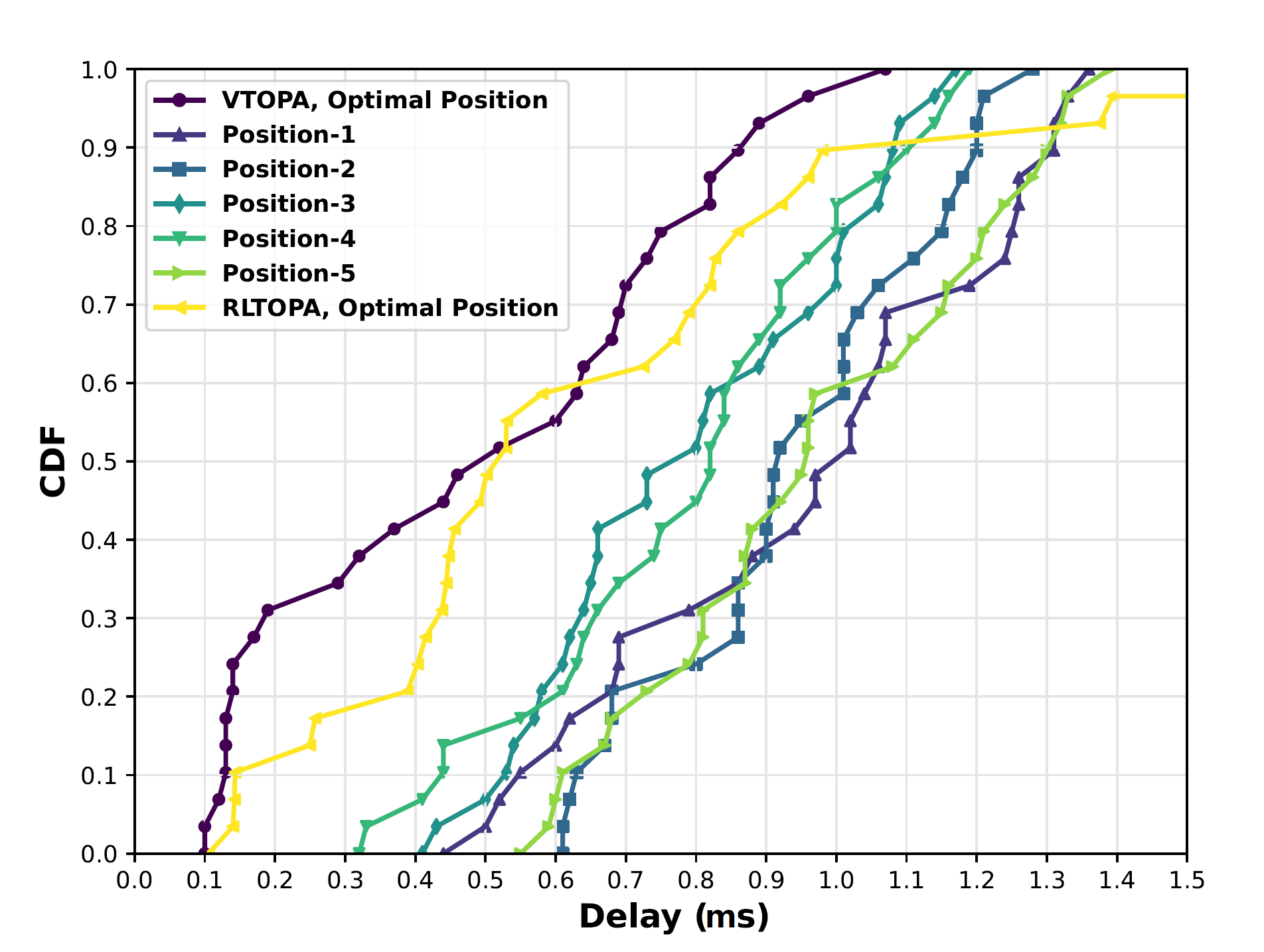}\label{fig9b}}
    \caption{Simulation results for Use Case A, comparing the performance of the UAV optimal position achieved by VTOPA with the optimal position selected by RLTOPA.}
    \label{fig9}
\end{figure}

\begin{figure*}[ht!]
	\centering
	\subfloat[Aggregate throughput (Mbit/s)] {
		\includegraphics[width=0.47\linewidth, height=6cm]{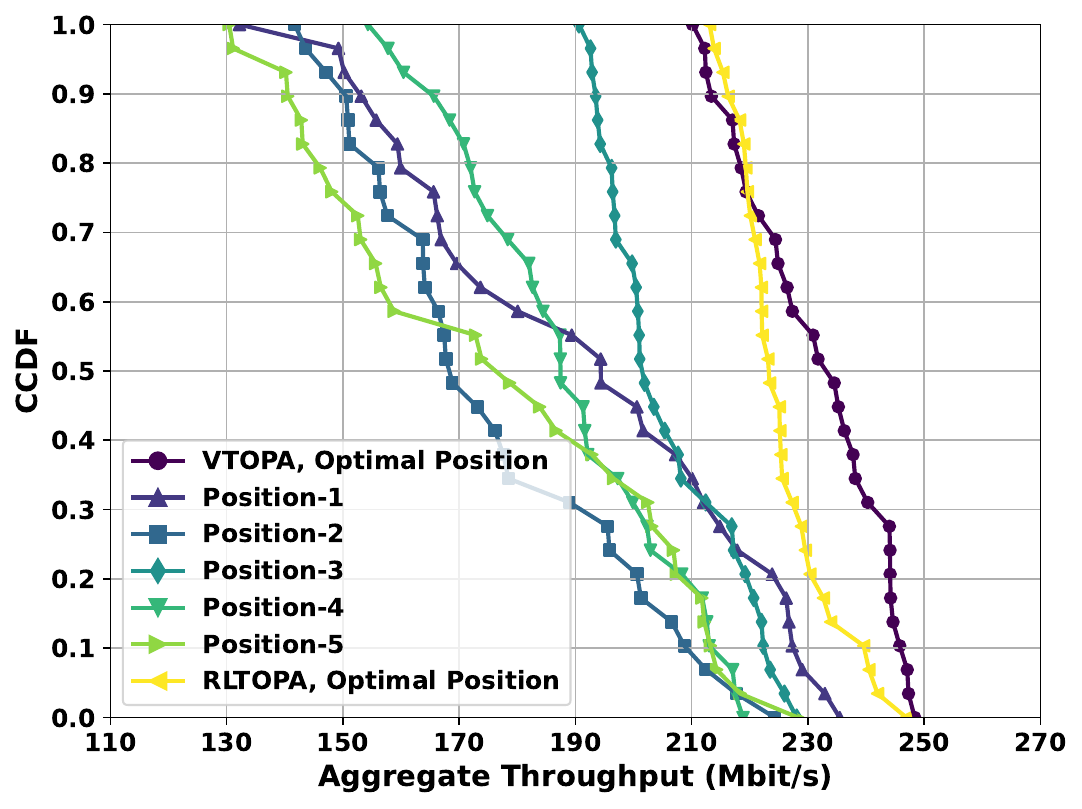}
		\label{fig10a}
	}
	\hfill
	\subfloat[Delay (ms)] {
		\includegraphics[width=0.47\linewidth, height=6.2cm]{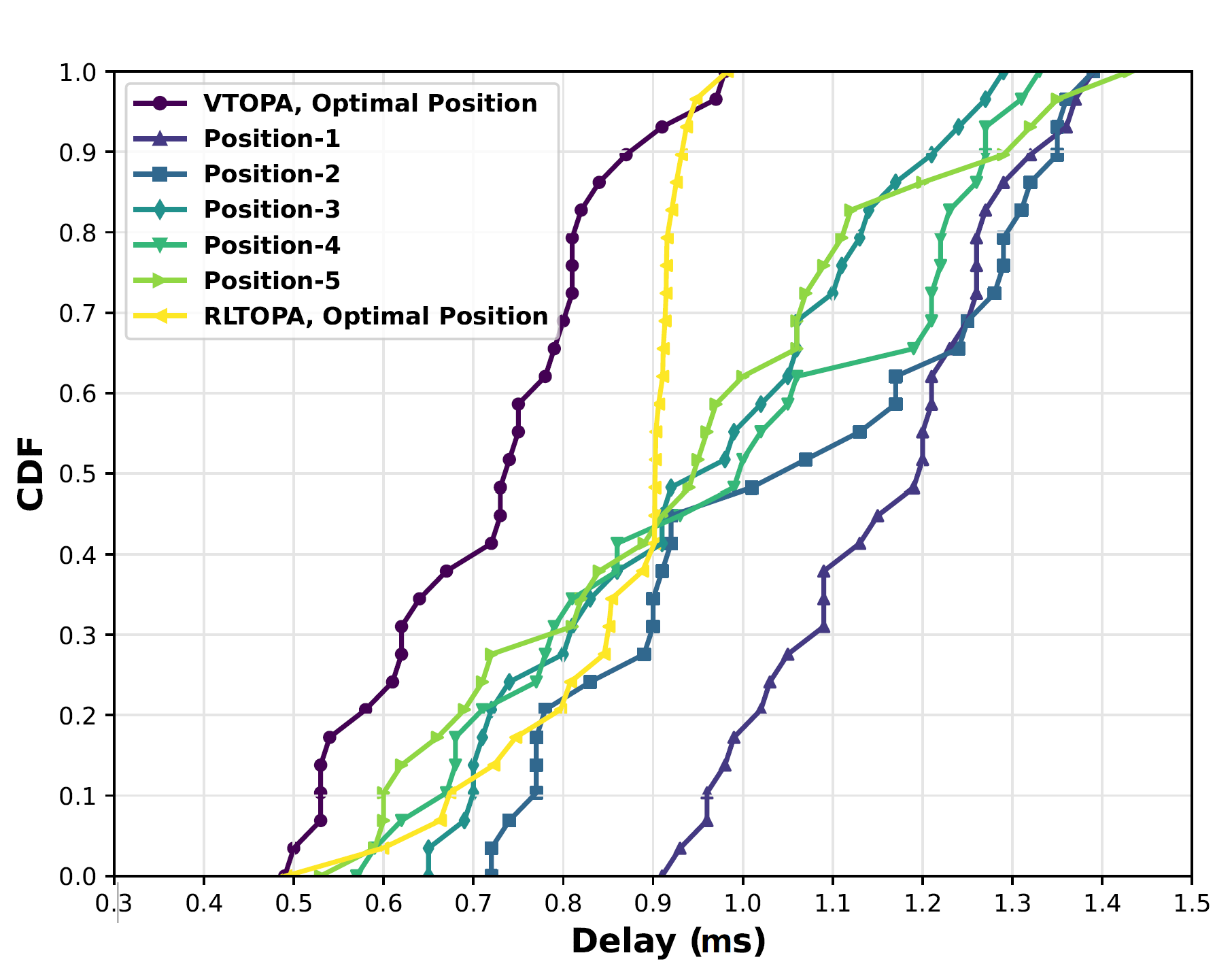}
		\label{fig10b}
	}
	\caption{Simulation results for Use Case B, comparing the performance of the UAV optimal position achieved by VTOPA with the optimal position selected by RLTOPA.}
	\label{fig10}
\end{figure*}

\begin{figure*}[ht!]
	\centering
	\subfloat[Aggregate throughput (Mbit/s)] {
		\includegraphics[width=0.47\linewidth, height=6cm]{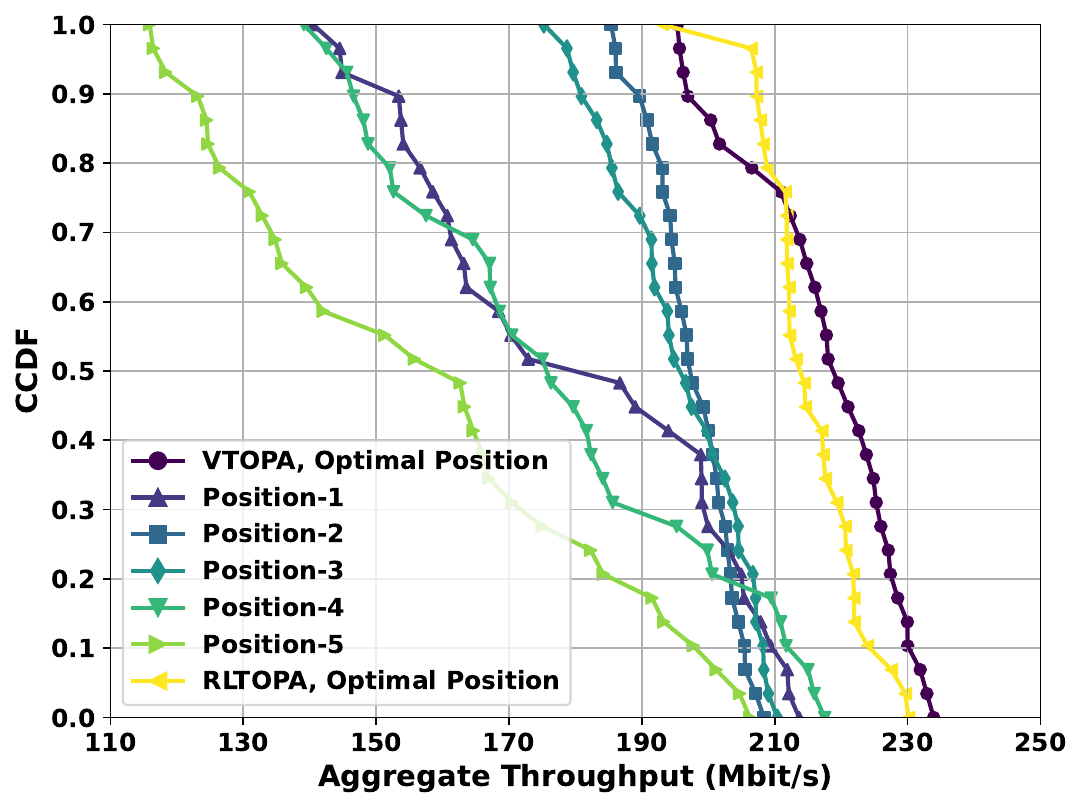}
		\label{fig11a}
	}
	\hfill
	\subfloat[Delay (ms)] {
		\includegraphics[width=0.47\linewidth, height=6.2cm]{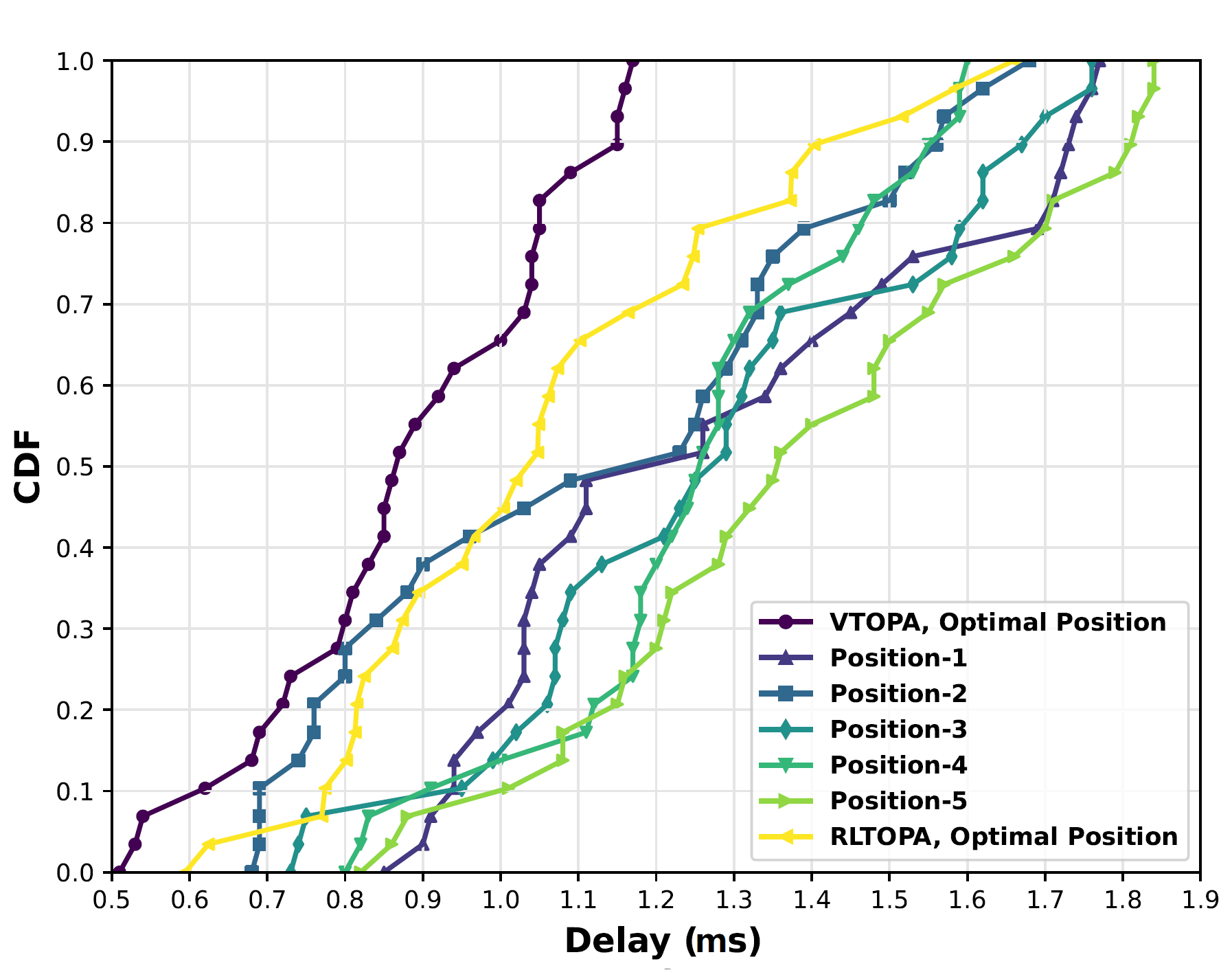}
		\label{fig11b}
	}
	\caption{Simulation results for Use Case C, comparing the performance of the UAV optimal position achieved by VTOPA with the optimal position selected by RLTOPA.}
	\label{fig11}
\end{figure*}
This section presents the simulation results. The performance of VTOPA is evaluated using Network Simulator 3 (ns-3) \cite{ns3adisc29:online}. A Wi-Fi Access Point (AP)-Station (STA) mode and a wireless connection were configured for the medium. All simulations were carried out under stable networking settings, considering the simulation parameters and the corresponding values presented in Table \ref{tab2}. The evaluation results were obtained from 30 simulation runs for each use case described in Section \ref{sec:Optimization Results}. The results are presented using the complementary cumulative distribution function (CCDF) to illustrate the distribution of aggregate throughput achieved by the UAV. The cumulative distribution function (CDF) is used to depict the mean delay. These results were obtained for the optimal positions determined by VTOPA. For comparison purposes, five alternative positions are tested in each case. These positions, labeled as \(Position-i\), where \(i \in \{1, ..., 5\}\), are located 10 meters away from the UAV optimal position in the left, right, front, back, and up directions, respectively.

\begin{table*}[t]
\footnotesize
\renewcommand{\arraystretch}{1.3}
\caption{Comparison between VTOPA and RLTOPA UAV positioning approaches}
\label{tab:pso_vs_rl}
\begin{tabular*}{\textwidth}{|l|p{0.299\textwidth}|p{0.3\textwidth}|}
\hline
\textbf{Metric} & \textbf{VTOPA (PSO)} & \textbf{RLTOPA (RL)} \\
\hline
Execution Time & A single run takes minutes (no training needed). \newline \scriptsize Main factors: \newline \scriptsize - number of iterations (100) \newline \scriptsize - number of particles (30) \newline \scriptsize - number of UEs to check & Requires hours/days for training. After training, execution is fast (seconds). \newline \scriptsize Training time depends on episodes (1000+) and complexity of environment \\
\hline
Memory Usage & Low memory usage. Only needs to store: \newline \scriptsize - List of candidate positions \newline \scriptsize - Current positions of particles \newline \scriptsize - Best positions found & High memory usage. Needs to store: \newline \scriptsize - Neural network model \newline \scriptsize - Training data \newline \scriptsize - Experience replay buffer \newline \scriptsize - State and action history \\
\hline
Convergence Speed & Fast convergence with iterative optimization & Requires training phase for policy learning \\
\hline
Exploration Strategy & Random position selection from valid set & Epsilon-greedy/policy-based exploration \\
\hline
Memory Requirements & Low (stores only particle positions) & High (stores Q-table/policy network) \\
\hline
Adaptability & High adaptability within defined search space & Can generalize to new scenarios \\
\hline
Position Discretization & Flexible discrete or continuous search space & Continuous or discretized action space \\
\hline
Computational Load & Low (simple position updates) & High (neural network computations) \\
\hline
Implementation Complexity & Simple (fewer hyperparameters) & More complex (requires tuning) \\
% \hline
% Multi-area Handling & Independent optimization per area & Can learn joint policy for all areas \\
\hline
\end{tabular*}
\end{table*}

Figure \ref{fig9} illustrates the performance of VTOPA for Use Case A. Figure \ref{fig9a} shows the aggregate throughput received by the UAV from the UEs. VTOPA achieves up to a 49\% improvement in throughput. For comparison purpose, we also evaluated in the same setting. As seen in the figures, both algorithms are efficient and significantly enhance network capacity. Figure \ref{fig9b} depicts the mean delay of transmissions toward the UAV, showing up to a 50\% reduction in delay. When compared to the optimal position achieved by RLTOPA, both algorithms demonstrate similar QoS improvements, particularly at the 50th percentile, where they exhibit comparable results.

Figure \ref{fig10} shows the evaluation results for Use Case B. Figure \ref{fig10a} demonstrates the effectiveness of both VTOPA and RLTOPA. Aggregate throughput improved by up to 38\% with VTOPA, while RLTOPA achieved an improvement of 32\%. In terms of mean delay (Figure \ref{fig10b}), VTOPA presents a greater reduction, decreasing the delay by up to 40\%, compared to RLTOPA's 25\% reduction. 

In Use Case C, the UAV is deployed within the intersection volume associated with the highest number of UEs. Figure \ref{fig11} depicts the performance for this use case. As shown in Figure \ref{fig11a}, both VTOPA and RLTOPA are efficient in this scenario; however, VTOPA presents greater improvement in throughput than RLTOPA, achieving up to a 38\% improvement. Figure \ref{fig11b} highlights the superior performance of VTOPA over RLTOPA in reducing delay. VTOPA offers a delay reduction of up to 35\%, whereas RLTOPA achieves a maximum reduction of 22\%.

Table \ref{tab:pso_vs_rl} compares VTOPA and RLTOPA with respect to several metrics, such as execution time, memory utilization, convergence characteristics, and implementation complexity. VTOPA leverages PSO to optimize UAV positions within a flexible search space, offering rapid and efficient solutions, requiring only minutes to execute without training. In contrast, the RL approach provides superior adaptability and continuous space operation at the expense of longer training times and higher computational demands. The primary difference lies in their core operational mechanisms: VTOPA achieves efficiency through iterative optimization in a flexible search space, while RL requires extensive training but excels in generalizing to novel contexts. VTOPA excels in scenarios requiring rapid deployment and computational efficiency, while RLTOPA is better suited for dynamic environments needing adaptability and generalization. This comparison highlights the trade-offs between immediate efficiency and long-term adaptability in UAV positioning tactics.

%%%%%%%%%%%%%%%%%%%%%%%%%%%%%%%%%%%%%%%%%%%%%%%%%%%%%%%%%%%%%%%%%
% Conclusions
%%%%%%%%%%%%%%%%%%%%%%%%%%%%%%%%%%%%%%%%%%%%%%%%%%%%%%%%%%%%%%%%%
\section{Conclusions} \label{Conclusions-Section}

This work introduced VTOPA, a vision-assisted UAV positioning strategy that integrates computer vision and wireless communication principles to address the challenge of optimal UAV placement in obstacle-rich scenarios. By considering visual information to extract environmental features and UE locations, VTOPA determines UAV positions that maximize LoS coverage while simultaneously accommodating the heterogeneous traffic demands. The proposed PSO-based solution was evaluated across multiple use cases and benchmarked against the previously developed reinforcement learning-based method, RLTOPA. Results show that while both approaches are effective, VTOPA consistently achieves superior performance in terms of execution time and positioning quality. For instance, VTOPA demonstrated throughput improvements of up to 50\% and latency reductions of up to 50\%, underscoring its suitability for real-time applications.

Nonetheless, the dual objective of maximizing LoS coverage and fulfilling all traffic demands introduces trade-offs, especially in dense user scenarios, as illustrated in Use Case C. These constraints highlight the complexity of UAV positioning under real-world conditions. Future work will focus on extending the framework to support multi-UAV deployments and dynamic scenarios with user mobility, enabling more scalable and adaptive network coverage solutions.

%%%%%%%%%%%%%%%%%%%%%%%%%%%%%%%%%%%%%%%%%%%%%%%%%%%%%%%%%%%%%%%%%
% REFERENCES
%%%%%%%%%%%%%%%%%%%%%%%%%%%%%%%%%%%%%%%%%%%%%%%%%%%%%%%%%%%%%%%%%
\bibliographystyle{elsarticle-num} 
\bibliography{References}

\end{document}